\documentclass[twocolumn, superscriptaddress, aps, nofootinbib]{revtex4}

\usepackage[utf8]{inputenc}
\usepackage[T1]{fontenc}
\usepackage[english]{babel} 
\usepackage{amsmath, amsfonts, amssymb, amsthm}
\usepackage{euscript}
\usepackage{tikz}
\usepackage{multirow}
\usepackage{url}
\usepackage{algorithm,algorithmic}

\newcommand{\C}{\mathbb{C}}

\DeclareMathOperator{\Log}{Log}
\DeclareMathOperator{\Tof}{Tof}
\DeclareMathOperator{\Sof}{Sof}
\DeclareMathOperator{\diam}{diam}
\DeclareMathOperator{\GG}{GG}
\DeclareMathOperator{\DFS}{DFS}
\DeclareMathOperator{\CE}{CE}

\newcommand\ket[1]{|#1\rangle}

\definecolor{ND}{rgb}{0,0,0} 
\newcommand\nicolas[1]{{\color{ND} #1}}

\begin{document}

\title{A Scalable Decoder Micro-architecture for Fault-Tolerant Quantum Computing}


\author{Poulami Das}
\affiliation{Georgia Institute of Technology, Atlanta, GA, USA}

\author{Christopher A. Pattison}
\affiliation{California Institute of Technology, Pasadena, CA, USA}

\author{Srilatha Manne}
\affiliation{Microsoft Quantum Systems and Microsoft Research, Redmond, WA, USA}

\author{Douglas Carmean}
\affiliation{Microsoft Quantum Systems and Microsoft Research, Redmond, WA, USA}

\author{Krysta Svore}
\affiliation{Microsoft Quantum Systems and Microsoft Research, Redmond, WA, USA}

\author{Moinuddin Qureshi}
\affiliation{Georgia Institute of Technology, Atlanta, GA, USA}

\author{Nicolas Delfosse\footnote{nidelfos@microsoft.com}}
\affiliation{Microsoft Quantum Systems and Microsoft Research, Redmond, WA, USA}

\begin{abstract}
\nicolas{
Quantum computation promises significant computational advantages 
over classical computation for some problems.  However, quantum 
hardware suffers from much higher error rates than in classical hardware.
As a result, extensive quantum error correction is required 
to execute a useful quantum algorithm.
The decoder is a key component of the error correction scheme whose
role is to identify errors faster than they accumulate in the quantum computer
and that must be implemented with minimum hardware resources
in order to scale to the regime of practical applications.
In this work, we consider surface code error correction, which is the most 
popular family of error correcting codes for quantum computing, 
and we design a decoder micro-architecture for the Union-Find decoding algorithm.
We propose a three-stage fully pipelined hardware implementation of the decoder
that significantly speeds up the decoder.
Then, we optimize the amount of decoding hardware required to perform 
error correction simultaneously over all the logical qubits of the quantum computer.
By sharing resources between logical qubits, we obtain a $67\%$ reduction 
of the number of hardware units and the memory capacity is reduced by $70\%$.
Moreover, we reduce the bandwidth required for the decoding process 
by a factor at least 30x using low-overhead compression algorithms.
Finally, we provide numerical evidence that our optimized micro-architecture 
can be executed fast enough to correct errors in a quantum computer.
}
\end{abstract}

\maketitle


Quantum computing promises significant speed-up over 
conventional computers for specific applications such as
integer factorization~\cite{shor1999polynomial}, 
physics and chemistry simulations~\cite{feynman,qchem, brown2010using}
and database search~\cite{grover1996fast}.

The primary obstacle to the implementation of quantum algorithms
solving industrial size problems is the high noise rate in any quantum
device that makes the output of a quantum computation indistinguishable 
from a random output.
A fault-tolerant quantum computer, in which quantum bits, or qubits,
are regularly refreshed by quantum error correction 
\cite{shorqec},
is necessary to perform a useful computation on noisy 
quantum hardware.

Most classical error correction schemes can be adapted to 
the quantum setting thanks to the CSS construction 
\cite{calderbank1996good, steanecode} 
and the stabilizer formalism 
\cite{gottesman1997stabilizer}, 
providing a quantum version of standard families of classical error-correcting codes \cite{macwilliams1977ECC} 
such as repetition codes, Hamming codes, Reed-Muller codes, BCH codes, LDPC codes 
and polar codes.

The main difference with the classical setting is the very high
noise rate that affects current quantum hardware, often of 
the order of $1\%$ per quantum gate, which makes quantum error correction 
much more challenging to implement.
This is because it relies on the measurement of quantum parity checks that 
are likely to introduce additional noise to the qubits.
Moreover, one must be able to implement a universal set of quantum gates 
on encoded qubits, or {\em logical qubits}.
In a {\em fault-tolerant quantum computer} (FTQC), the computation is performed
by alternating logical quantum gates and quantum error correction cycles 
that removes the noise injected by logical gates. 
Both quantum error correction and logical quantum gates must be 
implemented through {\em fault-tolerant gadgets} 
\cite{shor1996ft, aharonov1999ft_threshold, aliferis2005ft_threshold_d3}
to avoid the injection of pathological error configurations that would be 
uncorrectable by the subsequent correction cycle.

In this work, we consider a fault-tolerant quantum computer 
based on surface codes \cite{dennis2002tqm, fowler2012surface_code}
which is the most promising family of error-correcting codes for very noisy quantum hardware.
Surface codes can correct up to $1\%$ of noise on all the
basic components of the quantum computer and they 
can be implemented on a square grid of qubits using exclusively 
local quantum gates acting on nearest neighbor qubits.
For comparison, most quantum error correction codes such as quantum 
Hamming code (Steane code) requires an error rate below $10^{-5}$
\cite{svore2007steane_threshold_2d}.

In this paper, we focus on the design of an {\em error decoder}, or simply decoder, 
which is the primary building block in charge of error correction. 
The decoder takes as an input the {\em syndrome}, which is the data 
extracted from quantum parity check measurements and it returns an estimation
of the error. Given this estimation, the effect of the error can be easily reversed.
In a fault-tolerant quantum computer, the decoder must satisfy the three
following design constraints.
\begin{enumerate}
\item {\bf Accuracy Constraint:} 
The decoder must correctly identify the error with high probability.
\item {\bf Latency Constraint:} The decoder must be fast enough to correct errors 
within one error correction cycle.
\item {\bf Scalability Constraint:} The decoder design must feasible to implement 
in the regime of practical applications that may require millions of physical qubits.
\end{enumerate}
A number of surface code decoding algorithms have been proposed
in the past 20 years 
\cite{
dennis2002tqm,
fowler2012autotune,
fowler2012MWPM,
fowler2012MWPM_timing,
fowler2015parallel_MWPM,
duclos2013qudit_RG_decoder,
anwar2014qudit_SC_decoder,
watson2015qudit_decoder,
hutter2015qudit_improved_HDRG,
duclos2010RG_decoder,
bravyi2013RG_cubic_code,
duclos2013RG_improved,
barrett2010loss_correction_short,
stace2010loss_correction_long,
delfosse2017peeling_decoder,
fowler2013XZ_correlations,
delfosse2014XZ_correlations,
criger2018XZ_correlations_and_degen,
tomita2014LUT_decoder,
heim2016optimal_decoder,
dennis2005phd,
harrington2004phd,
wootton2012high_threshold_decoder,
hutter2014MCMC_decoder,
wootton2015simple_decoder,
herold2015FT_SC_decoder,
breuckmann2016local_decoder_2d_4d,
delfosse2017UF_decoder,
kubica2019CA_decoder,
nickerson2019correlation_adaptive_decoder,
ferris2014TN_QEC,
bravyi2014MLD,
darmawan2018Lineartime_TN_decoding,
tuckett2018TN_biased,
torlai2017NN_SC_decoder,
baireuther2018NN_SC_correlations,
krastanov2017DNN_SC_decoder,
varsamopoulos2017NN_SC_decoding,
chamberland2018DNN_small_codes,
maskara2019NN_CC_decoder,
breuckmann2018NN_high_dimension_codes,
sweke2018RL_SC_decoder,
baireuther2019NN_CC_circuit_decoder,
ni2018NN_SC_large_d,
andreasson2019DNN_toric_code,
davaasuren2018CNN_SC_decoder,
liu2019NN_LDPC_codes,
varsamopoulos2018NN_SC_decoder_design,
varsamopoulos2019NN_SC_decoder_distributed,
wagner2019NN_training_with_symmetry,
chinni2019NN_CC_decoder,
sheth2019NN_SC_combining_decoder,
colomer2019RL_TC_decoder}
and many satisfy the Accuracy Constraint which is
often the main motivation.
However, it remains unclear whether the decoding can be made fast
enough to satisfy the Latency Constraint without degrading the decoder 
performance \cite{fowler2017QEC_talk}.
Moreover, the decoding problem is generally studied for a single 
logical qubit, ignoring the Scalability Constraint, whereas practical applications 
require hundreds or thousand of logical qubits encoded into millions of physical qubits
\cite{qchem}.
A substantial amount of decoding hardware is required to decoding simultaneously 
all the qubits of the quantum computer.
Most of the work in quantum error correction decoders focuses
on algorithmic aspects of the problem. Here, we consider this problem 
through the lens of computer architecture and we propose a decoder 
micro-architecture that satisfies simultaneously our three design contraints. 

\begin{figure}[htb]
\centering
    \includegraphics[width=.9\columnwidth]{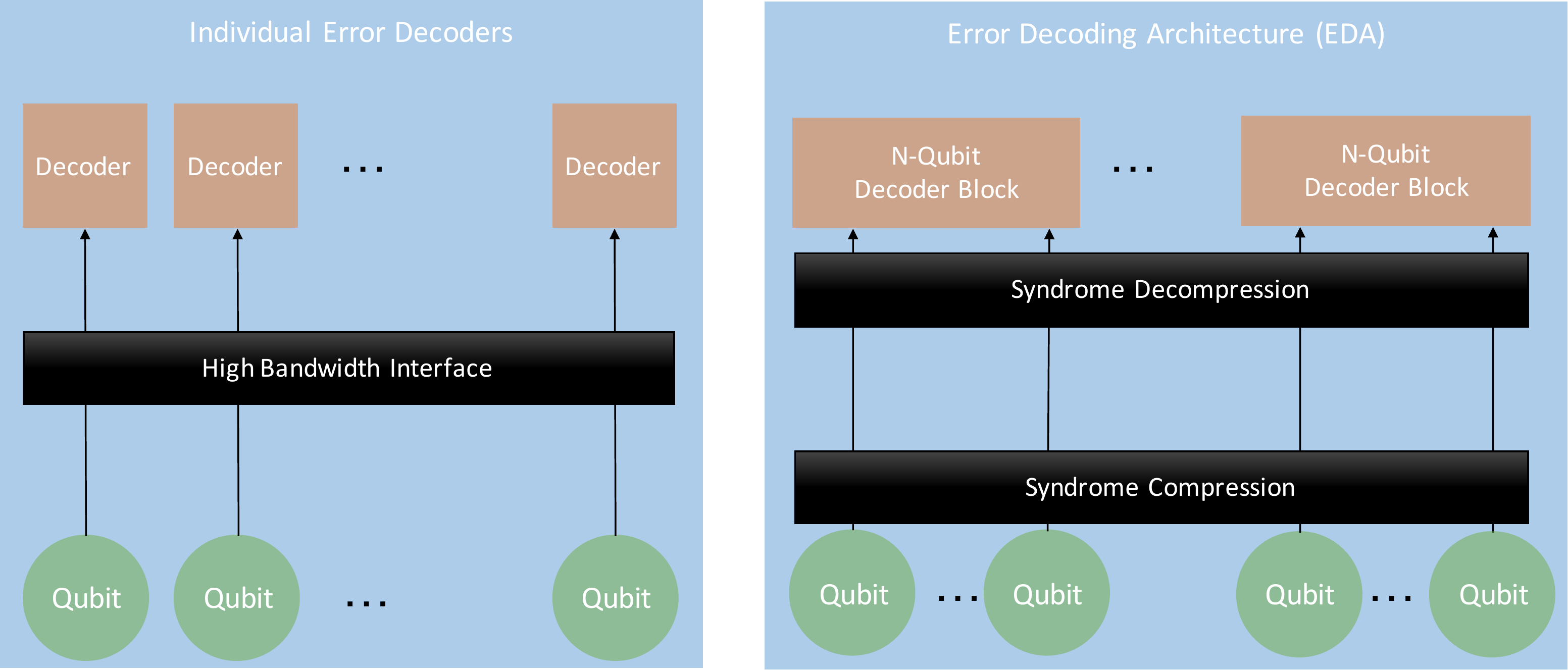}
    \caption{ A naive decoding architecture with a decoder associated with each 
    logical qubit and our Error Decoding Architecture based on optimized decoder
    blocks shared across multiple logical qubits. A low-overhead compression algorithm
    is used to reduce the bandwidth cost. Our optimized decoder block is described 
    in Fig.~\ref{fig:dblock}.
	}
    \label{fig:EDA}
\end{figure}

We propose a decoder micro-architecture based on the 
Union-Find decoding algorithm \cite{delfosse2017UF_decoder, delfosse2017peeling_decoder}.
We choose the Union-Find decoder for its accuracy and its simplicity.
It is proven to achieve good decoding performance.
Moreover, it comes with a almost-linear time complexity and 
it is also very fast in practice because it requires no floating-point 
arithmetic and no matrix operations.
Finally, the simplicity of the decoder allows us to design a specialized
hardware implementation that leads to a significant speed-up.
Our main contributions are the following.
\begin{enumerate}
\item We propose a hardware implementation of the Union-Find decoder based on 
three hardware units corresponding to each step of the algorithm.

\item We design a three-stage pipepline based on our three hardware units
that brings an important speed-up to the Union-Find decoder by 
parallelizing the implementation of the decoding stages.

\item We observe that the utilization of different components 
of the Union-Find decoder pipeline varies with the unit type and across
the logical qubits and hence propose an efficient time-division 
multiplexing that allows logical qubits to share decoding resources 
within a decoder block without compromising the error correction capabilities.

\item We propose different compression algorithms adapted to a cryogenic 
setting in order to reduce the bandwidth consumed to send the syndrome
data to the decoder. 

\item Combining all the previous results, we describe an 
{\em Error Decoding Architecture } (EDA) represented in Fig.~\ref{fig:EDA}
that scales the decoder design for a large FTQC while reducing 
hardware costs.
The number of hardware units is reduced by $67\%$ and we obtain 
a $30 \times$ bandwidth reduction while preserving the decoder accuracy.
Moreover, we demonstrate by numerical simulations that our architecture 
leads to a decoder that is fast enough to satisfy the Latency Constraint, 
despite the high noise rate of quantum hardware.
\end{enumerate}

Item 1 and 2 provide a hardware acceleration of the decoder and
3 and 4 lead to a satisfying solution to the Scalability Constraint.
Our EDA is optimized carefully in order to guarantee that the 
error correction capability of the initial UF decoder is preserved, which
guarantees that the Accuracy Constraint is satisfied.
The numerical simulation of our optimized micro-architecture (item 5) 
accounts for the limitation imposed by shared hardware resources
which demonstrates that our decoder satisfy simultaneously the three design 
constraints.

This article exploits a number of ideas from computer architecture such as
pipelining and resource optimization. We believe this approach is necessary 
in order to scale quantum machines. Even though we provide a detailed
micro-architecture for the Union-Find decoder, the general principles of our
design apply to any decoding algorithm.
A key ingredient is the simplicity of the decoder and the decomposition
in independent steps, which leads to a natural pipeline and a speed-up
by instruction-level parallelism.

\section{Background and Motivation}
\label{sec:bg}
\subsection{Qubits and decoherence}

\nicolas{
We refer to Preskill's lecture notes \cite{preskill1998lecture}
and Nielsen and Chuang's book \cite{nielsen2002quantum} 
for a great overview of field of quantum computation and quantum information 
theory.
A qubit is the basic unit of information in a quantum computer. 
A qubit is described by complex vector $\ket \psi = \alpha \ket{0} + \beta \ket{1}$, 
which represents the superposition of two basis state $\ket 0$ and $\ket 1$, with 
$\alpha, \beta \in \C$ such that $|\alpha|^2 + |\beta|^2 = 1$.
Without error correction, a quantum state rapidly {\em decoheres} 
due to the accumulation of tiny rotations.
By constantly measuring the system, one can project these tiny rotations onto
three types of errors denoted $X, Y$ and $Z$ and called {\em Pauli errors}.
The bit-flip error $X$ swaps the basis state $\ket 0$ and $\ket 1$
and maps the qubit $\ket \psi$ into $\beta \ket 0 + \alpha \ket 1$.
The phase-flip error $Z$ is defined by $Z \ket \psi = \alpha \ket 0 - \beta \ket 1$.
It introduces a relative phase between the two basis states.
The error $Y$ corresponds to a simultaneous bit-flip and phase-flip error, {\em i.e.}, $Y = XZ$ 
up to an overall phase which does not affect the result of the computation.
We use the notation $I$ for the identity operation that corresponds to the
error-free case.

By definition of $Y$, it is enough to correct $X$-type and $Z$-type errors.
In this work, we focus on the correction of $X$-type errors. By symmetry of the 
error correction scheme consider in this article, $Z$-type errors can be corrected 
using the exact same mechanisms by swapping the roles of $X$ and $Z$.
For simplicity, we assume that the correction of $X$-type and $Z$-type errors 
is performed independently, although a performance gain is possible by taking into account
the correlations between $X$ and $Z$.
\cite{
fowler2013correlation, 
delfosse2014XZcorrelations, 
tuckett2018ultrahigh}.
}

\subsection{Surface codes}

\nicolas{
In order to combat decoherence, we must perform the computation on 
{\em encoded data}, also called {\em logical data}, which is corrected at 
regular time intervals using a quantum error correcting code 
\cite{
shorqec,
steanecode,
calderbank1996good, 
gottesman1997stabilizer}. 
In this work, we focus on the surface code 
\cite{
dennis2002topological,
fowler2012surface_code}
which is the most promising quantum error correcting code for a 
quantum computing architecture due to its high error threshold, 
which means that the error correction protocol can be implemented 
with very noisy quantum hardware.
An error rate below $1\%$ for all the components of the quantum computer is
sufficient in order to obtain encoded qubits with better quality than our
initial physical qubits \cite{
raussendorf2007fault, 
raussendorf2007topological, 
fowler2009highthreshold}.
In order to scale to the massive sizes required for practical applications, 
it is necessary to build quantum hardware whose fault rate is far 
below the $1\%$-threshold. This is because error correction 
does not decrease the error rate sufficiently if the initial
qubit error rate is too close to the threshold.
In this work, we consider a noise rate of $10^{-3}$.
}

\begin{figure}[htb]
\centering
    \includegraphics[width=.9\columnwidth]{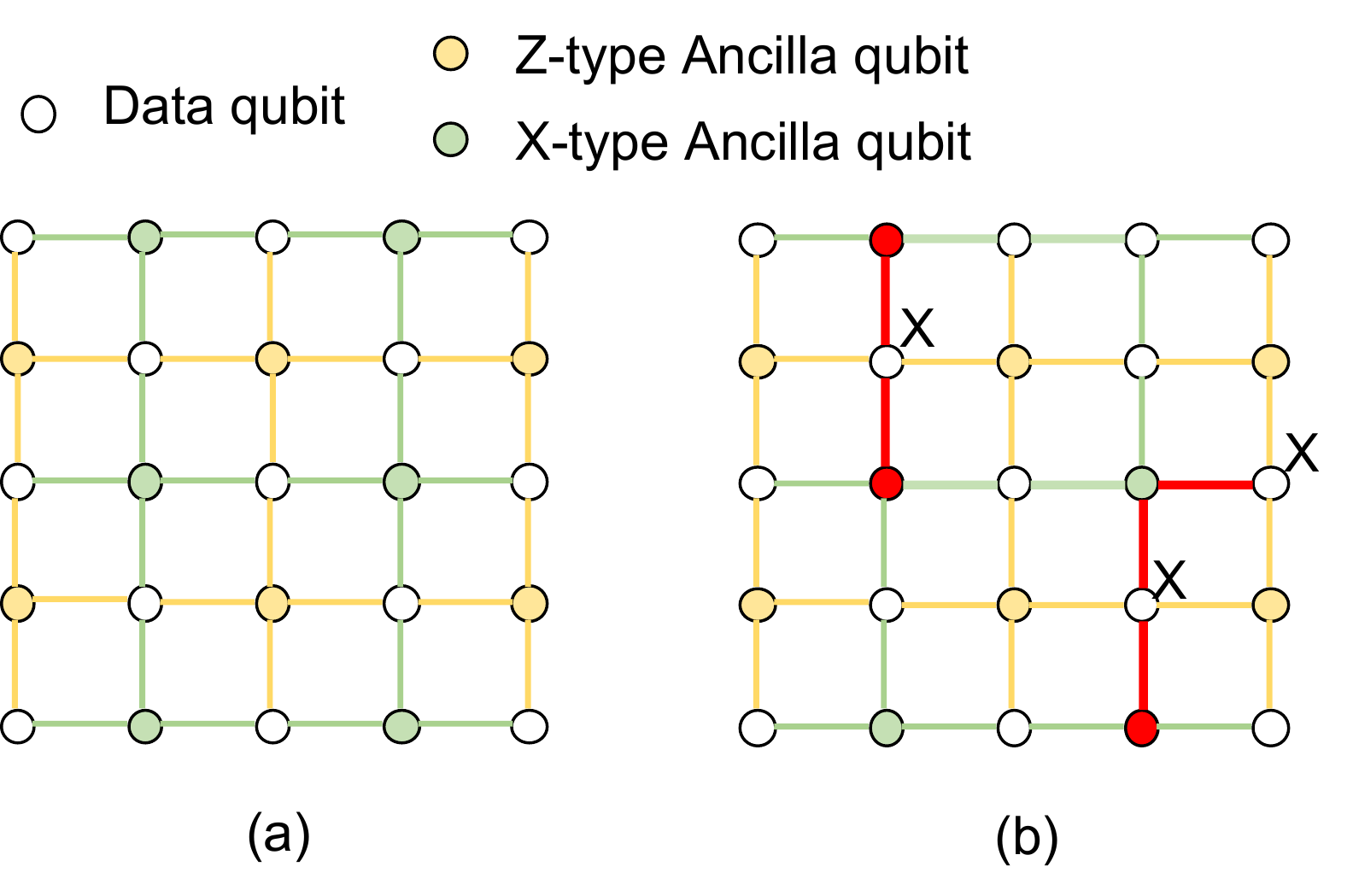}
    \caption{(a) Distance-three surface code SC(3). Data qubits store the logical information and $X$-type (resp. $Z$-type) ancilla qubits are used to extract the syndrome of $X$-type (resp. $Z$-type) errors.
    (b) A set of $X$-errors detected by non-trivial syndrome values on red nodes.
	}
    \label{fig:surfacecode}
\end{figure}

The family of surface codes is the most widely considered candidate for
designing a fault-tolerant quantum computer. 
The distance-$d$ surface code, that we denote SC($d$), 
encodes a logical qubit into a square grid of $(2d-1) \times (2d-1)$ qubits, 
alternating {\em data qubits}, which store the logical information, 
and {\em ancilla qubits}, used to detect errors as shown in 
Figure~\ref{fig:surfacecode} with the distance-three surface code SC(3).
The main reason for the success of the surface code is its locality which 
significantly simplifies the quantum chip design.
Error correction with surface codes only requires local interactions between ancilla
qubits and their nearest neighbor data qubits, that is at most four qubits. 
The minimum distance $d$ of the code measures the error correction capability.
A larger minimum distance $d$ results in an increased error tolerance at the 
price of an increased qubit overhead.

\nicolas{
Encoding physical qubits that suffer from an error rate $p$ using a
distance-$d$ surface code, we obtain a logical qubits with error rate
\begin{align} \label{eq:logical_error_rate}
p_{\Log}(d, p) = 0.15 \cdot (40 \cdot p)^{(d+1)/2}
\end{align}
that we call {\em logical error rate}. 
This heuristic formula, derived from the numerical results of 
\cite{delfosse2017unionfind}, provides a good approximation 
of the logical error rate in the regime of low error rate ($p << 10^{-2}$).
It is valid in the context of the current work, that is when using
the Union-Find decoder and for the phenomenological noise model
introduced in Section~\ref{subsec:decoding_problem}.

Throughout this article, we illustrate our design with numerical results for 
distance-11 surface codes which is a reasonable distance for a first generation 
of fault-tolerant quantum computer since it allows to implement non-trivial
quantum algorithms on logical qubits while keeping the qubit overhead 
to a few hundred qubits per logical qubit. Assuming a physical error rate
of $p = 10^{-3}$, the logical qubits error rate drops to 
$
p_{\Log} \approx 6 \cdot 10^{-10}
$ 
allowing for the implementation of large depth quantum algorithms.
}

\subsection{The decoding problem} \label{subsec:decoding_problem}

\nicolas{
In this section, we review the decoding problem and the graphical formalism 
introduced in \cite{dennis2002topological}.
}

Quantum error correction is a two-step process. 
First, a measurement circuit is executed producing a syndrome 
bit for each ancilla qubit. All syndrome bits can be extracted simultaneously.
Then, a {\em decoding algorithm} is used to identify errors on data qubits 
based on the syndrome information. 
To avoid any confusion with other decoding operations used in this architecture,
we sometimes refer to the decoder as the error decoder.
\nicolas{
The decoding subroutine is a purely classical operation that can be 
performed on highly reliable hardware. On the other hand, 
the syndrome extraction is implemented on noisy quantum hardware.
In order to obtain sufficiently accurate information about the errors occurring on
data qubits despite measurement errors, multiple rounds of syndrome extraction 
are performed.
The decoder analyzes $d$ consecutive rounds of syndrome data to produce
an estimation of the error induced on the data qubits.
}

\begin{figure}[htb]
\centering
    \includegraphics[width=.9\columnwidth]{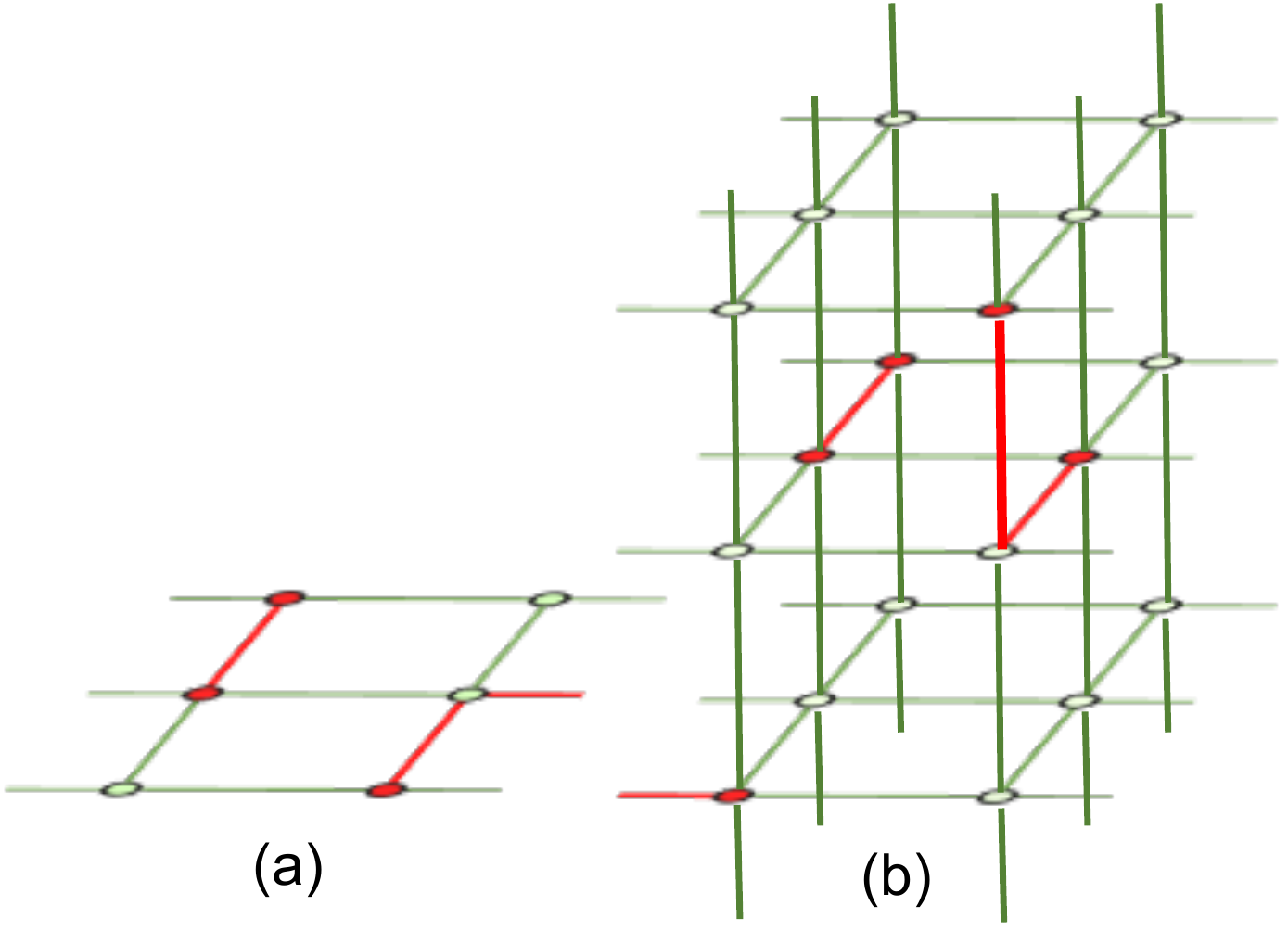}
    \caption{(a) The planar decoding graph corresponding to a single round of syndrome
    extraction without measurement error. 
    Qubits are supported on edges and syndrome bits correspond to vertices.
    Red edges mark the presence of $X$-type errors on data qubits
    and red nodes show the non-trivial syndrome bits extracted.
    (b) To fight against measurement errors, three rounds of syndrome extraction
    are performed, represented by three layers of the graph (a) connected
    by vertical edges corresponding to measurement errors.
    Errors are represented by red edges and the endpoints of an error path 
    is detected by a non-trivial syndrome (red node). 
    No syndrome bit is extracted on the left and right boundaries of the lattice
    nor on the bottom and top boundary. 
	}
    \label{fig:decoding_graph}
\end{figure}

\nicolas{
In the absence of measurement errors, the decoding problem can be 
mapped onto a matching problem in a square grid as shown in Fig.~\ref{fig:decoding_graph}.
}
A bit-flip error $X$ acting on a single data qubit is detected by a non-trivial 
syndrome bit on the incident $X$-type ancilla qubits as Fig.~\ref{fig:surfacecode}(b)
shows.
More generally, a chain of $X$-errors is detected by its endpoints.
In order to recover the chain of $X$-errors given the syndrome values (its endpoints), 
the basic idea of the decoding algorithm is to build a short chain of $X$-type errors 
that matches the detected endpoints. 
\nicolas{
Errors on the boundary of the lattice are detected only on one of their endpoints.
}

\nicolas{
To address the issue of measurement errors, $d$ rounds of syndrome extraction
are performed, resulting in a matching problem in a three-dimensional 
graph \cite{dennis2002topological}.
Fig.~\ref{fig:decoding_graph} shows the cubic graph 
representing errors and syndromes bits for the distance-3 surface code.
In what follows, we refer to this graph as the {\em decoding graph}.

We simulate errors occurring on data qubits and incorrect syndrome values
using the phenomenological noise model \cite{dennis2002topological}.
Each edge of the decoding graph corresponds to a potential error,
with horizontal edges representing $X$-errors on data qubits and vertical
edges corresponding to syndrome bit-flips. 
We assume that an error occurs on each edge, independently, with probability $p$. 
For each vertex $v$ in the bulk of the lattice, a syndrome value $s(v) \in \{0, 1\}$ 
is extracted, which is the parity of the number of errors incident to $v$.
Just like in the planar case, the goal of the decoder is to estimate the error by 
matching together the vertices $v$ supporting a non-trivial syndrome
$s(v) = 1$. 
This formalism allows to handle both qubit errors and measurement errors
in the same way.

The relevance of the phenomenological noise model is justified in 
\cite{dennis2002topological}.
For further study of our decoding architecture tailored to a specific type
of qubit, one may consider a more precise noise model such as the 
circuit-level noise model \cite{dennis2002topological}.
In this work, we chose the phenomenological noise model because it 
is simple enough to develop a good intuition about the decoder and 
it captures the essential properties of the quantum hardware which 
guarantees that all the ideas proposed in the current work generalize to 
a more precise model and remain relevant for a practical device.
}

\subsection{Existing Error Decoding Algorithms}
\label{sec:backgroundondecoders}

In this section, we discuss different decoding strategies.
The noise model describes the probability of all possible errors. Given this information, 
we can derive the probability of each error when a given syndrome is observed. 
The ultimate goal of the decoder is to return  an error whose probability is maximal
given the syndrome measured, {\em i.e.}, a most likely error. 
A decoder that achieves this performance is said to be a maximum likelihood decoder~\cite{dennis2002topological}.
For an arbitrary error-correcting code, it is generally not possible to 
implement efficiently a maximum likelihood decoder~\cite{hsieh2011np}.
However in the case of the surface code, 
several algorithms achieve a good error correction performance.
In what follows, we discuss several promising decoding strategies.

{\bf Lookup Table (LUT) decoder}~\cite{tomita2014low}: 
This decoder implements a maximum likelihood decoder 
using a lookup table indexed by the syndrome bits. 
The corresponding LUT entry stores the correction to be applied to the data qubits. 
The LUT size grows exponentially with code distance making this design unsuitable 
for large FTQCs. 

\textbf{Minimum Weight Perfect Matching (MWPM) decoder}~\cite{dennis2002tqm}: 
The MWPM decoder provides an estimation of a most likely error based on a graph pairing algorithm, 
the MWPM algorithm, that can be implemented in polynomial time \cite{kolmogorov2009blossom}.
This decoder is one of the most effective in terms of error correction capabilities, 
even though its worst-case time complexity, $O(|V|^3) \propto O(d^9)$ where $|V|$ is the size of the decoding graph, makes it too slow for large-distance surface codes.
Fowler suggested a parallel implementation of this algorithm that reduces the average 
time complexity to $O(1)$, although the worst case complexity remains significant \cite{fowler2015parallel_MWPM}.
This decoder relies on large amounts of parallelism from several ASICs 
for each logical qubits but this study does not discuss the system architecture or 
the number of ASICs needed.

\textbf{Machine Learning (ML) decoder}: 
ML-decoders train neural networks with the underlying error probability 
distribution and decoding is then treated as an \textit{inference} problem 
where the syndrome data is an input to the neural network which infers 
the correction~
\cite{
torlai2017NN_SC_decoder,
baireuther2018NN_SC_correlations,
krastanov2017DNN_SC_decoder,
varsamopoulos2017NN_SC_decoding,
chamberland2018DNN_small_codes,
maskara2019NN_CC_decoder,
breuckmann2018NN_high_dimension_codes,
sweke2018RL_SC_decoder,
baireuther2019NN_CC_circuit_decoder,
ni2018NN_SC_large_d,
andreasson2019DNN_toric_code,
davaasuren2018CNN_SC_decoder,
liu2019NN_LDPC_codes,
varsamopoulos2018NN_SC_decoder_design,
varsamopoulos2019NN_SC_decoder_distributed,
wagner2019NN_training_with_symmetry,
chinni2019NN_CC_decoder,
sheth2019NN_SC_combining_decoder,
colomer2019RL_TC_decoder}
ML-Decoders require substantial computational resources and the size of the training 
data grows quickly with the code distance. They also require large training times, 
and are primarily studied for small code distances. 
There exists some preliminary studies for larger code distances~\cite{breuckmann2018NN_high_dimension_codes, ni2018NN_SC_large_d} 
and proposals to obtain better performance through distributed neural networks~\cite{varsamopoulos2019NN_SC_decoder_distributed}, 
and hardware platforms~\cite{breuckmann2018NN_high_dimension_codes} 
such as GPUs, FPGAs, and TPU~\cite{jouppi2017datacenter}.

{\bf Tensor Network  (TN) decoder}
\cite{ferris2014TN_QEC,
bravyi2014MLD,
darmawan2018Lineartime_TN_decoding, 
tuckett2018TN_biased}
The probability of each possible error can be represented as a tensor network
which leads to a decoding algorithm that contracts this tensor network. 
The contraction of the tensor network requires extensive matrix operations which 
may be hard to scale, however the algorithm achieves a very good error-correction 
performance that is optimal \cite{bravyi2014MLD} or quasi-optimal.
Although it has not been studied precisely, tensor network decoders 
could benefit from a hardware speed-up from neural accelerators such as a TPU~\cite{jouppi2017datacenter}.

\textbf{Union-Find (UF) decoder}~\cite{delfosse2017unionfind, delfosse2017peeling}: 
This is a recently proposed algorithm that offers a correction in 
almost-linear time $O(n \alpha(n))$, where $\alpha(n)$ is $\leq 5$ 
for all practical purposes. 
It uses Union-Find data structure in order to achieve a similar performance 
as the MWPM decoder without using computationally intensive matching algorithms.

\begin{table}[htb]
\begin{center}
\caption{Abstract comparison of decoder accuracy, latency and scalability (adapted from ~\cite{varsamopoulos2019NN_SC_decoder_distributed})}
\setlength{\tabcolsep}{1.2mm} 
\renewcommand{\arraystretch}{1.0}
\label{tab:decodercomparison}
\begin{footnotesize}
\begin{tabular}{ |c|c|c|c| } 
\hline

Decoder & Acuracy & Latency & Scalability\\
\hline
LUT 
	& Very High 
	& Low 
	& Poor \\
\hline
TN 
	& Very high 
	& Moderate 
	& Moderate  \\
\hline
MWPM 
	& High to Very high
	& Moderate
	& Moderate \\
\hline
ML 
	& High
	& Low 
	& Moderate \\
\hline
UF 
	& High 
	& Low 
	& High \\
\hline
\end{tabular}
\end{footnotesize}
\end{center}
\end{table}

Table~\ref{tab:decodercomparison} provides an abstract comparison of 
prominent decoding algorithms in terms of the three key design constraints 
highlighted in introduction: accuracy, runtine and scalability.

Given the simplicity of the algorithm and low time-complexity, we use the 
UF decoder as the default algorithm for our studies.
However, the design principles, optimizations, and scalability analysis of the
present work will hold true for other decoders as well. 
Similarly, the syndrome compression analysis in 
Section~\ref{sec:compression} applies for any decoder-quantum substrate 
interface irrespective of the decoder and qubit technology in use.

\subsection{Union-Find decoder}
\label{sec:edalgos}

In this section, we review the strategy of the 
Union-Find decoder \cite{delfosse2017unionfind, delfosse2017peeling}.
As explained in the introduction, our principal motivation for choosing
this decoding algorithm is its rapidity and its simplicity.

\begin{figure}[htb]
\centering
    \includegraphics[width=.9\columnwidth]{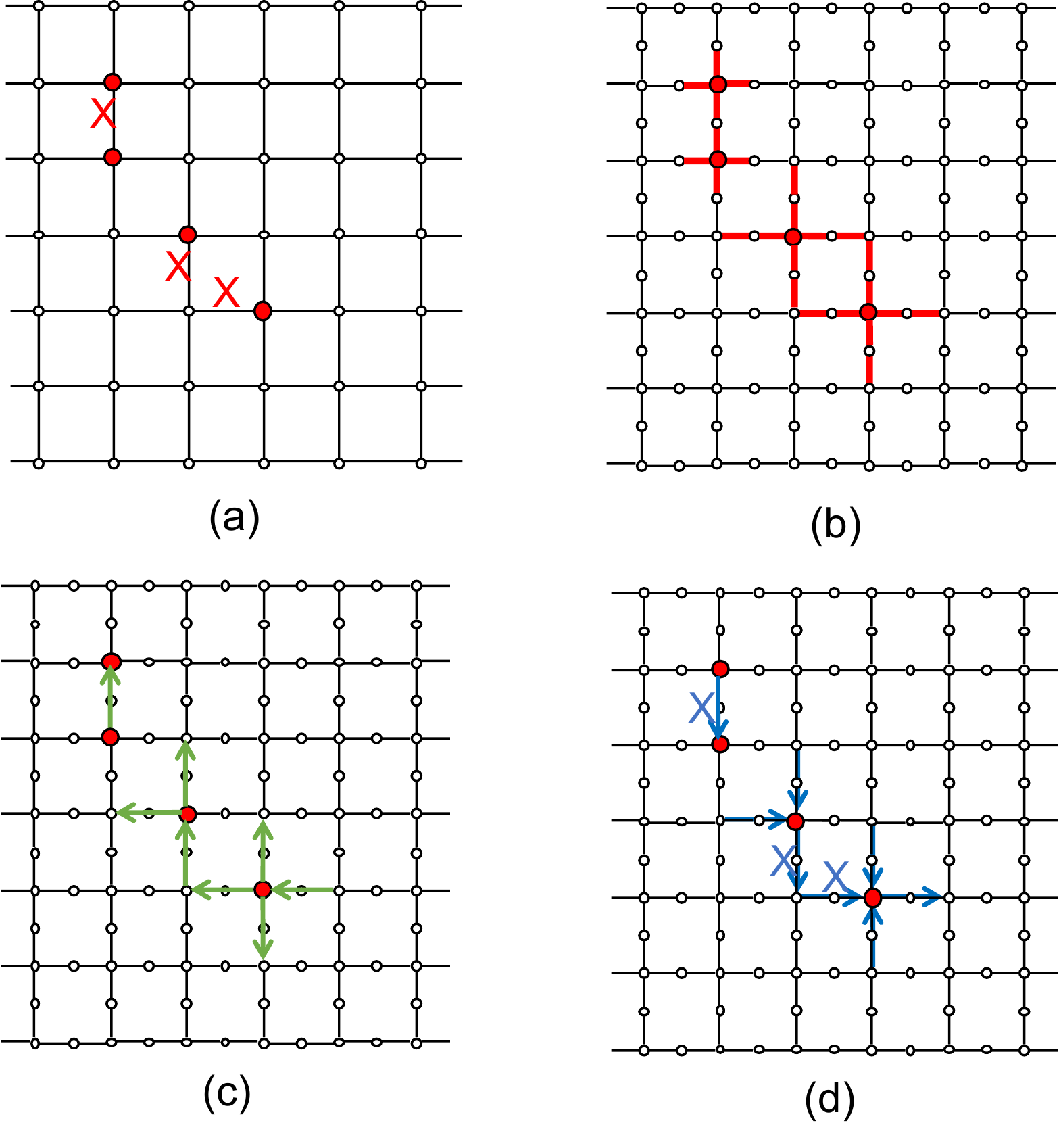}
    \caption{The three main steps of the Union-Find decoder for a 2d decoding
    graph with the surface code SC(7).
    (a) A $X$-type error and its syndrome in the decoding graph. The goal is to recover the error
    given the red syndrome nodes. 
    To mark half-edges we will add a vertex in the middle of each edge of the decoding
    graph.
    (b)~{\bf Cluster Growth:} We keep growing clusters around the red 
    nodes by adding half-edges in all directions.
    The growth of a cluster stops when it contains an even number of red nodes
    or if it meets the boundary. The top cluster grows only one step.
    The bottom cluster requires two growth steps.
    (c)~{\bf Spanning Tree:} Build a spanning tree for each grown cluster 
    in the decoding graph. Ignore half-edges.
    (d)~{\bf Peeling:} Identify the error on the edges of the spanning trees from the leaves
    to the root.
	}
    \label{fig:UF_steps}
\end{figure}

The Union-Find decoder operates in three steps that we review in Figure~\ref{fig:UF_steps}.
We illustrate the decoding procedure using a two-dimensional decoding graph
since there is no major difference with the cubic case that are relevant in practice.
The algorithm takes as an input a set of nodes supporting non-trivial
syndrome values and its goal is to recover the error living on the edges of the
decoding graph.
(a) During the first step, even clusters are grown around non-trivial syndrome nodes.
(b) Then, a spanning tree is built for each cluster and is oriented from a root to the leaves. 
(c) Finally, we build an estimation of the error using the syndrome by traversing the clusters
in reverse order.

\section{Hardware design and pipelining}

\begin{figure}[!h]
\centering
    \includegraphics[width=1.0\columnwidth]{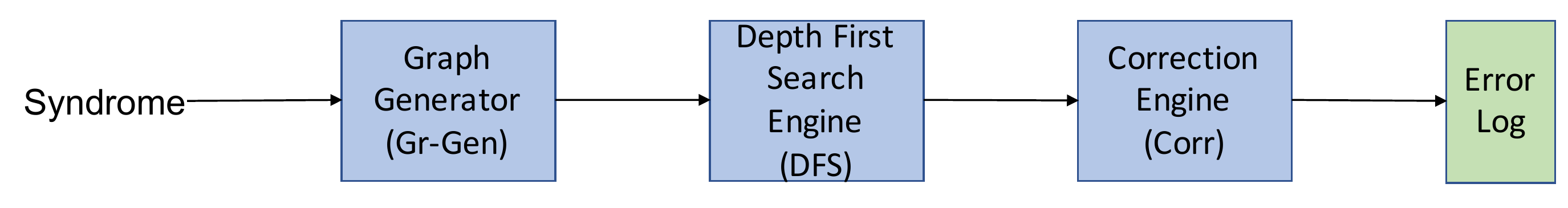}
    \caption{Block Diagram of UF decoder pipeline with three units corresponding to the
	three steps of the UF decoding algorithm.}
    \label{fig:blockdiag}
\end{figure}

\nicolas{
In this section, we design three hardware units implementing the three stages
of the Union-Find decoder as Figure~\ref{fig:blockdiag} shows.
The Graph Generator (Gr-Gen) produces the grown clusters obtained 
is Figure~\ref{fig:UF_steps}(b).
The Depth First Search (DFS) engine generates the spanning forest 
from Figure~\ref{fig:UF_steps}(c).
The correction is implemented as in Figure~\ref{fig:UF_steps}(d)
by the Correction (Corr) engine.
}

To reduce the bandwidth and latency issues it is more favorable to 
operate the decoding circuitry close to the quantum substrate
in a cryogenic regime.
Our main design constraint is the limited hardware resources available 
in a cold environment. We optimize our design to minimize the memory
requirement and the number of memory reads.

\nicolas{
Our implementation of the decoding algorithm benefits from hardware acceleration in two ways. 
First, a fully pipelined design allows performance improvement through enhanced parallelism across the different processing units. While the correction engine works on cluster $i$, the DFS engine can build the spanning tree for cluster $i-1$.
Second, in a general purpose processor, the read latency depends upon where the data is present and it can range up to several hundreds of cycles if it needs to be fetched from the off-chip main memory to the on-chip caches. For our specialized hardware, the processing elements can directly access the data stored on-chip that require much fewer cycles.
In this work, we assume a readout time of four cycles to read 32-bit data.
}

\label{sec:ufdesign}

\subsection{Graph Generator}

The Graph Generator module takes the syndrome as an input
and generates a spanning forest by growing clusters around 
non-trivial syndrome bits (non zero syndrome bits). Instead 
of growing all surrounding edges as in Figure~\ref{fig:UF_steps}(b)
we only add the edges that reach new vertices.
This directly produces a spanning forest without extra cost.
The spanning forest is built using two fundamental graph operations: 
\textit{Union()} and \textit{Find()} \cite{tarjan1975UF}. 

\begin{figure}[!t]
\centering
    \includegraphics[width=1.0\columnwidth]{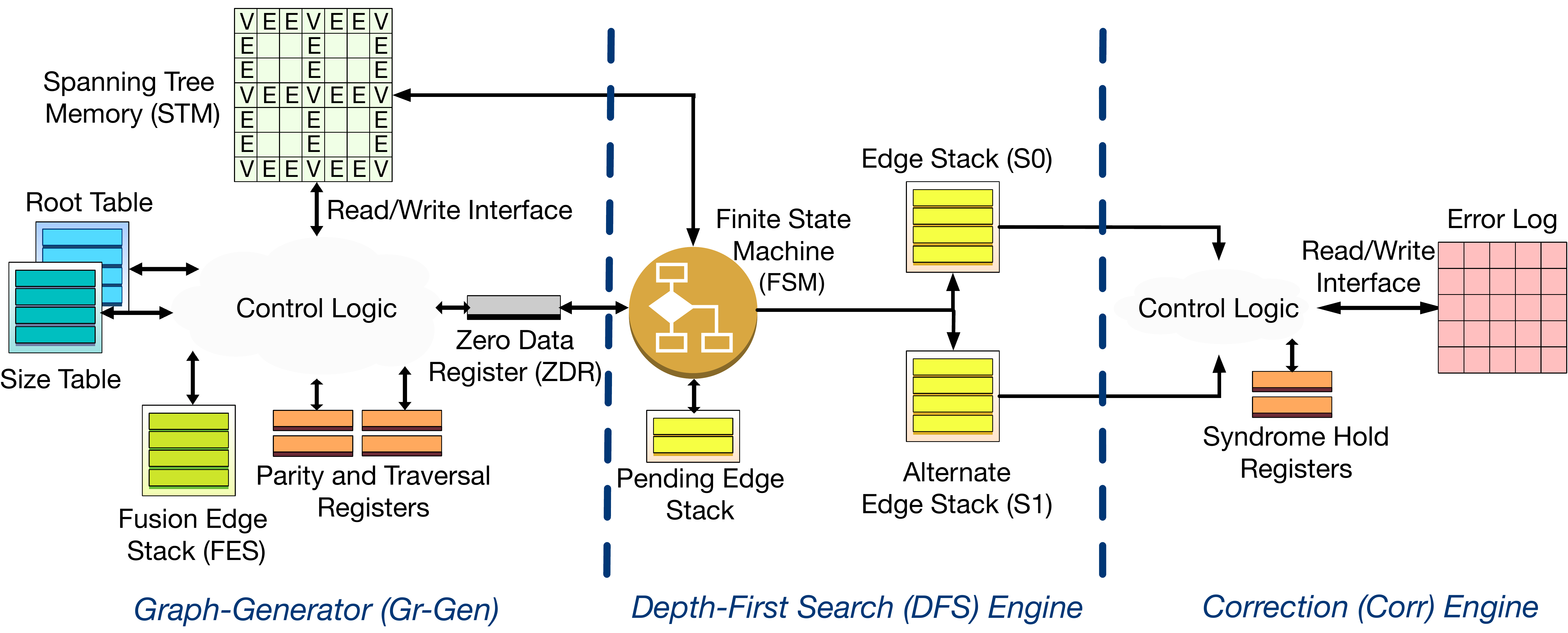}
    \caption{Block Diagram for the whole decoding pipepline including the three decoding modules.}
    \label{fig:pipelinedatapath}
\end{figure}

Figure~\ref{fig:pipelinedatapath} shows the design of the
three modules that implements the decoding algorithm. 
The Gr-Gen module consists of the Spanning Tree Memory (STM), 
a Zero Data Register (ZDR), a root table, a size table, parity registers, 
and a fusion edge stack (FES) 
\footnote{Our proposed design is slightly different from the actual 
UF algorithm since our objective is not obtain achieve an optimal
asymptotic complexity but to reduce the cost of hardware resources
for a given system size}.
The size of each component is a function of the code distance $d$. 
The STM stores one bit for each vertex, and two bits per edge. 
Two bits per edge are required since clusters grow around a 
vertex or existing cluster boundary by half edge width as per the 
original decoding algorithm. 
The ZDR indicates whether a row of the STM contains all zeros 
by storing a "0" for rows where all bits are 0, and a "1" for rows where 
there are one or more non-zero bits. 
Since the total number of edges in the spanning forest are generally small, 
the ZDR accelerates the STM traversal. 
The fusion edge stack (FES) stores the newly grown edges so that 
they can be added to existing clusters. The root table and size table 
stores the root and size of a cluster, respectively.

\begin{figure}[!b]
    \centering
    \includegraphics[width=1.0\columnwidth]{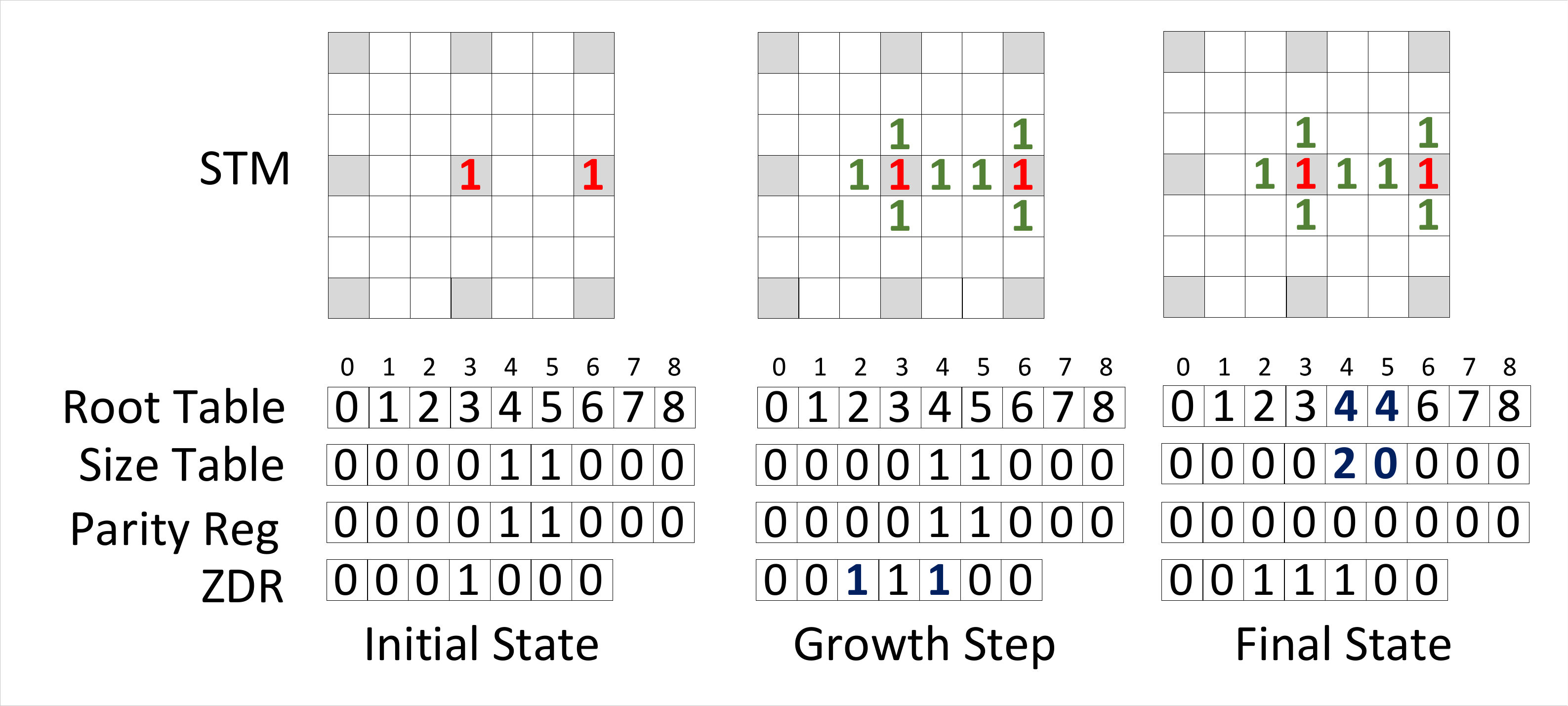}
    \caption{State of the major components in Gr-Gen module during in graph generation}
    \label{fig:grsteps}
    \end{figure}

The root table entries are initialized to the indices RootTable[i] = i, as shown 
in Figure~\ref{fig:grsteps}. 
The size table entries for the non-trivial syndrome bits are initialized to 
1 as shown in Figure~\ref{fig:grsteps}. 
These tables aid the \textit{Union()} and \textit{Find()} operations to merge clusters after the growth phase. They are indexed by cluster indices. The tables are sized for the maximum number of clusters possible which is equal to the total number of vertices in the surface code lattice. The tree traversal registers store the vertices of each cluster visited in the \textit{Find()} operation. Since the decoding algorithm grows all odd clusters until the parity is even, odd clusters must be detected quickly. To do the same, we use parity registers as shown in Figure~\ref{fig:pipelinedatapath}. The parity registers store 1 bit parity per cluster depending upon whether it is odd or even. For a reasonable code distance of 11, seven 32-bit registers are enough. For larger code distances, we store the additional parity information in the memory and read them in advance when required to hide the memory latency.

    \begin{figure}[!h]
    \centering
    \includegraphics[width=1.0\columnwidth]{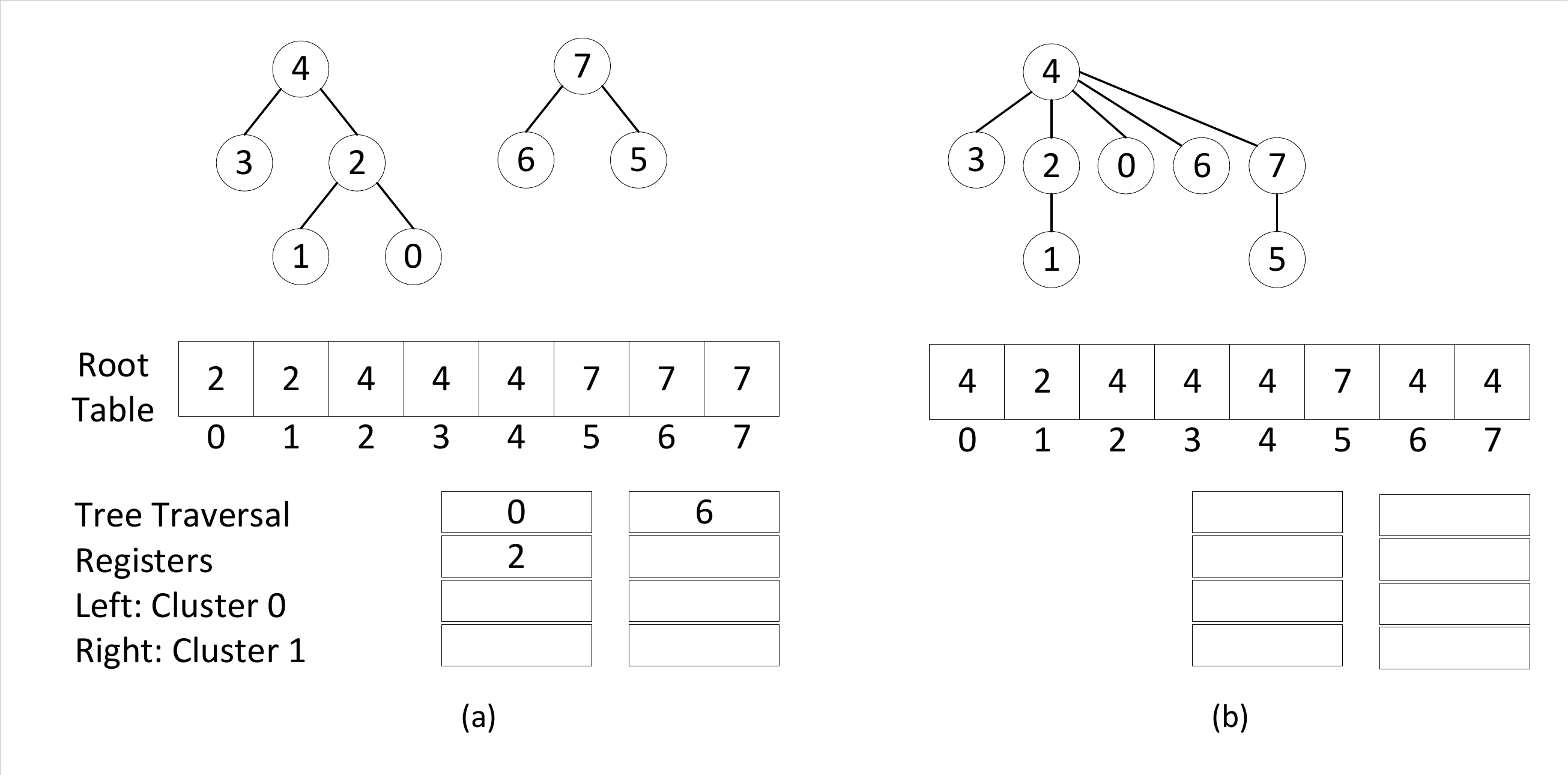}
    \caption{ (a) State of two clusters and root table entries. Primary vertices 0 and 6 are now connected by an edge and the two clusters needs to be merged. Vertices encountered on the \textit{Find()} path are saved on the tree traversal registers (b) Root table entries updated for vertices encountered on the \textit{Find()} path.}
    \label{fig:ttrs}
    \end{figure}

The control logic reads the parity registers and grows clusters with odd parity (called the growth phase) by writing to the STM, ZDR, and adding newly added edges that touches other cluster boundaries to the FES. The STM is not updated for edges that connect to other clusters to prevent double growth. It is updated when clusters are merged by reading from the FES. The logic checks if a newly added edge connects two clusters by reading the root table entries of the vertices connected by the edge (call these the primary vertices). This is equivalent to the \textit{Find()} operation. The vertices visited on the path to find the root of each primary vertex are stored on the tree traversal registers as shown in Figure~\ref{fig:ttrs}(a). The root table entries for these vertices are updated to directly point to the root of the cluster to minimize the depth of the tree for future traversals. This operation, called \textit{path compression}, is a key feature of the Union Find algorithm and enables the reduction of the tree depth, amortizing the cost of \textit{Find()} operation. For example, Figure~\ref{fig:ttrs}(a) shows the state of two clusters and root table at an instant in time. Let us assume that after a growth step, vertices 0 and 6 are connected and the two clusters must be merged. The tree traversal registers are used to update the root of vertex 0 as shown in Figure~\ref{fig:ttrs}(b). Since the depth of the tree is compressed during every \textit{Find()} operation, only a few 32-bit registers are sufficient. The proposed design uses 5 registers per primary vertex. If the primary vertices belong to different clusters, the root of the smaller cluster is updated to point to the root of the larger cluster.

Delfosse et. al. store the boundary of each cluster in their 
algorithm~\cite{delfosse2017unionfind}. 
Based on a Monte-Carlo simulation that shows that the average 
cluster diameter remains very small in the noise regime 
that is relevant for practical applications, we decided to compute the cluster 
boundary when it is needed in the growth phase, 
instead of consuming extra memory to store it
\footnote{
By computing the boundary indices during cluster growth, 
we reduce the required memory capacity by \textbf{10\%} at the cost of 4 adders in the Gr-Gen.}.

To summarize, the Gr-Gen module detects odd parity clusters using the parity registers and grows then by reading and writing to the STM. The cluster growth is aided by the information stored on the root table, size table and FES.

\subsection{Depth First Search Engine}

The DFS engine processes the STM data produced by the Gr-Gen that stores the set of grown even clusters. It uses the depth first search algorithm to generate the list of edges that forms a spanning tree for each cluster in the STM\footnote{A breadth first search exploration works too but we prefer DFS since it is generally more memory efficient}. 
The logic is implemented using a finite state machine and two stacks as shown in Figure~\ref{fig:pipelinedatapath}. 
Stacks are used since the order in which edges are visited in the spanning tree must be reversed to perform correction by peeling~\cite{delfosse2017peeling}. The edge stack stores the list of visited edges while the pending edge stack is used as to queue the next edges to explore in the on-going DFS.

To enable pipelining and improve performance, we design the micro-architecture to consist of an alternate edge stack  (Edge Stack 1 as shown in Figure~\ref{fig:pipelinedatapath}). When there is more than one cluster, the correction engine works on the edge list of one of the traversed clusters when the DFS engine traverses through the other. The DFS Engine generates the list of edges visited to traverse a cluster using DFS algorithm and hence the number of memory reads required is directly proportional to the size of the clusters. By going over the STM row-wise and using the ZDR to visit only non-zero rows, the effective cost of generating clusters is reduced. 

\begin{figure}[htb]
\centering
    \includegraphics[width=1.0\columnwidth]{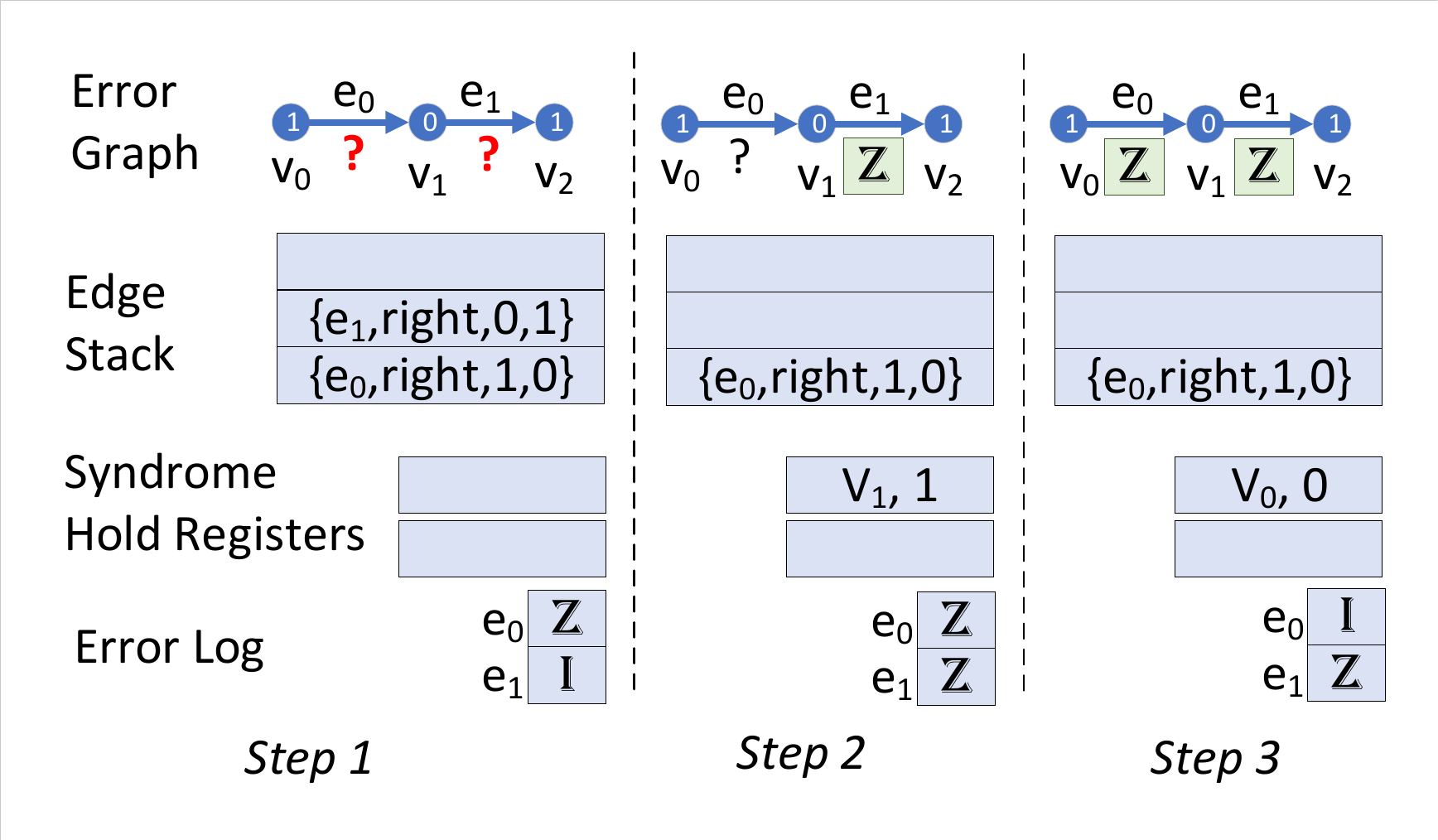}
    \caption{Peeling for an example error graph performed in the Correction Engine. The status of the edge stack, error log, and syndrome hold register are shown for each step.}
    \label{fig:correctionengine}
\end{figure}

\subsection{Correction Engine} 

The correction engine performs the peeling process of the decoder~\cite{delfosse2017peeling} and identifies the Pauli correction to apply. This requires access to the edge list (which is stored on the stack) and syndrome bits corresponding to the vertices along the edge list. The syndrome bits can be accessed by accessing the STM. However, this increases the logical complexity, latency, and the number of memory requests that the STM is required to handle. To reduce the incoming memory traffic for the STM and eliminate the need for additional logic, the syndrome information is saved on the stack along with the edge index information by the DFS Engine. The temporary syndrome changes caused by peeling are saved on local registers (Syndrome Hold Registers shown in Figure~\ref{fig:correctionengine}). The Corr Engine also reads the last surface code cycle error log and updates the Pauli correction for the current edge. For example, if the error on a edge $e_0$ was $Z$ in the previous logical cycle and it encounters a $Z$ error in the current cycle too, the Pauli error for $e_0$ is updated to $I$ as shown in Figure~\ref{fig:correctionengine}.

\subsection{Hardware cost} \label{subsec:HW_design:HW_cost}

We measure the hardware cost by estimating the amount of memory required.
Table~\ref{tab:memory_req} shows the different contributions to the memory
requirement.

The spanning tree memory (STM) used by the Gr-Gen and DFS engine accounts 
for most of the storage costs.
It contains 1 bit per node of the decoding graph and at most 2 bits per edge (only 1 bit
on the boundary).
The decoding graph is a 3d cubic lattice with about $d^3$ vertices which leads to
a total of $7 d^3$ bits for the STM.

The root table and the size table used in the Gr-Gen module contains
$d^3$ entries and each entry consists of an integer index
which can be uniquely identified using $\log_2(d^3)$ bits. 
Thus, the total sizes of the root and size tables are $3 d^3 \log_2(d)$ each.


The size of the edge stacks $S_0$ and $S_1$ used by the DFS and the Corr engine 
is given by the maximum number of edges of a spanning tree of a cluster.
The size of the spanning tree of a cluster $C_i$ is given by 
$|V(C_i)| - 1$ where $|V(C_i)|$ denotes the number of vertices of the cluster $C_i$.
To fit any possible cluster, one could pick a stack that can store $d^3$ edges,
which requires about $d^3 \log_2(d^3)$ bits.

For simplicity, we ignore the Fusion Edge Stack and the Pending Edge Stack that 
are in general significantly smaller the two edge stacks $S_0$ and $S_1$.
This is because these stacks contain only a small subset of edges of a cluster.

\begin{table}[htb]
\begin{center}
\begin{footnotesize}
\caption{Memory requirement for our hardware design of the Union-Find decoder as a function 
of the minimum distance $d$ of the surface code.
Note that we consider the worst case but the two Edge Stacks can be made significantly smaller as explained in Section~\ref{subsec:HW_design:HW_cost}.}
\setlength{\tabcolsep}{1.2mm} 
\renewcommand{\arraystretch}{1.0}
\label{tab:memory_req}
\begin{tabular}{ |c|c|c|c|c| } 
\hline 
	& size in bits
 	& $d=5$
	& $d=15$
	& $d=25$
	\\
\hline 
	STM 
	& $7 d^3$
	& 100 Bytes
	& 3 KBytes
	& 14 KBytes
	\\
\hline
	Tables ($\times 2$)
	& $3 d^3 \log_2(d)$
	& 100 Bytes
	& 5 KBytes
	& 27 KBytes
	\\
\hline
	Parity Reg.
	& $d^3$
	& 15 Bytes
	& 400 Bytes
	& 2 KBytes
	\\
\hline
	ZDR
	& $3d^3$
	& 46 Bytes
	& 1.3 KBytes
	& 6 KBytes
	\\
\hline
	Edge stacks ($\times 2$)
	& $3 d^3 \log_2(d)$
	& 100 Bytes
	& 5 KBytes
	& 27 KBytes
	\\	
\hline
\end{tabular}
\end{footnotesize}
\end{center}
\end{table}

\begin{figure}[htb]
\centering
    \includegraphics[width=0.9\columnwidth]{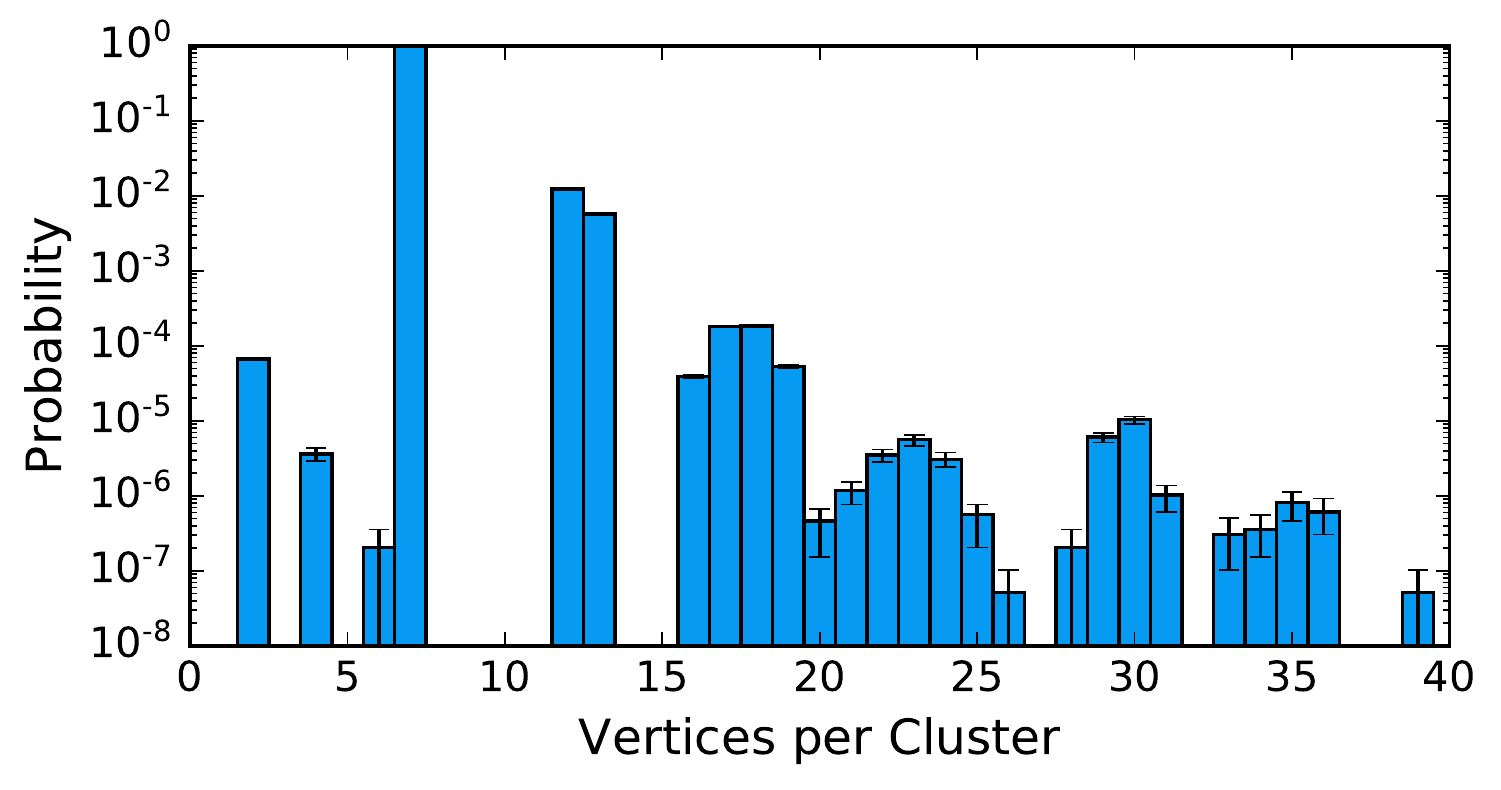}
    \vspace{-0.15in}
    \caption{Distribution of the number of vertices per cluster after the growth step of the 
    UF decoder for code distance $d=11$ and physical error rate $p=10^{-3}$.
    This distribution is estimated by a Monte-Carlo simulation using $10^7$ samples.
	}
    \label{fig:edgedist}
\end{figure} 

\nicolas{
The size of the edge stacks can in principle reach the previous upper 
bound $d^3 \log_2(d^3)$, in the case of a cluster that covers the whole 
decoding graph.
This makes it the most expensive element of the design. 
However, the probability that such a large cluster is reached
remains extremely small in a practical noise regime
\footnote{This is because a large cluster generally correspond to an uncorrectable 
error configuration and by definition, the probability of such an error
is at most $p_{\log}$.}.
By analyzing the maximum size of a cluster after the growth step 
using Monte Carlo simulations, we optimize the stack size for a 
chosen code distance $d$ and a given physical error rate $p$.
Figure~\ref{fig:edgedist} shows the cluster size distribution
for code distance $d = 11$ and physical error rate $p=10^{-3}$.  
For these parameters, the probability of a cluster of size 
larger than 80 is smaller than the logical error rate for this code. 
One can thus ignore the clusters of size larger than 80 without 
significantly affecting error correction performance.
This drops the stack memory requirement by a factor 10x 
from 1.7 KBytes to 0.13 KBytes.

To reduce further the memory requirement, the edge stacks can be
sized to half the maximum size of a cluster spanning tree (derived from simulation), 
optimizing for common case. 
In the rare event of an overflow, the alternate stack is used.

In general, the memory required for each of the two DFS stacks 
is approximately $3 S(d, p) \log_2(d)$
where 
$S(d, p)$ is the minimum integer $s$ such that 
the probability to have a cluster with more than $s$ edges 
on the output of the Gr-Gen is at below 
$
p_{\Log}(d, p)
$. 
We say that a {\em stack overflow failure} occurs if the DFS
engine encounters a cluster that does not fit in a stack, that is
with more than $S(d, p)$ edges.
By construction, the stack sizes are optimize in such a way 
\begin{align} \label{eqn:stack_overflow_failure}
p_{\Sof}(d, p) \leq p_{\Log}(d, p)
\end{align}
where $p_{\Sof}$ denote the probability of a decoding failure 
due to a stack overflow error.
}

\subsection{Comparison with other decoders}

For comparison, we provide a rough estimate of the memory capacity required for
the MWPM decoder.
The average number of faults in the decoding graph for a given noise rate
is $p |E|$ where $|E|$ is the number of edges of the decoding graph.
For low values of $p$ (such as $p = 10^{-3}$), 
many of these configurations of $p |E|$ faulty edges are non-overlapping,
and most of the edges are in the bulk of the lattice,
which results in about $w = \lceil 2p |E| \rceil$ non-zero syndrome bits.
Given a set of $w$ non-zero syndrome nodes, the MWPM generates 
a complete graph with $w+1$ vertices 
(and $w(w+1)/2$ edges) and it performs a Minimum Weight Matching 
algorithm in this graph.
A state of the art implementation of this algorithm due to 
Kolmogorov \cite{kolmogorov2009blossom}
consumes about 161 bits per edge (4 pointers, 1 integer and 1 bit per edge)
which brings the memory capacity for the MWPM decoder to at least 
\begin{align} \label{eq:MWPM_mem_bound1}
161 \cdot \lceil 2p |E| \rceil \cdot ( \lceil 2p |E| \rceil + 1) / 2 \approx 2900 \cdot p^2 d^6
\end{align}
bits.
Therein, we use $|E| \approx 3 d^3$.
In order to provide a non-trivial lower bound when $d$ is small, we
use the fact that the MWPM decoder must correct at any set 
of $(d-1)/2$ faults. Such a set may lead to $w = (d-1)$ non-zero syndrome bits, 
resulting in a lower bound of $161 (d-1) d / 2 \approx  80 d^2$ bits.

\begin{figure}[htb]
\centering
  \includegraphics[width=0.90\linewidth]{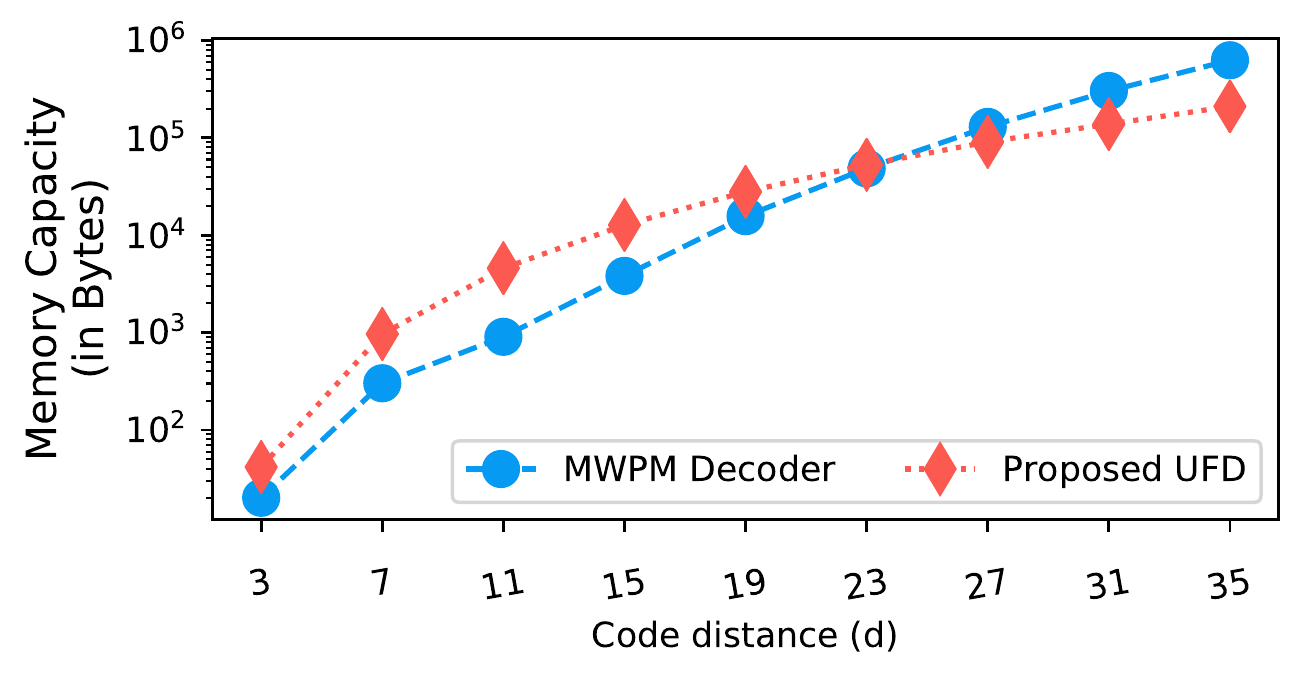}
\vspace{-0.1in}
\caption{Comparison of memory capacity required for the MWPM decoder and for our implementation of the UF decoder.}
\label{fig:uf_vs_mwpm_memory}
\vspace{-0.1in}
\end{figure}

Taking the best of both lower bounds, we obtain the result of 
Figure~\ref{fig:uf_vs_mwpm_memory} which shows that our 
UF decoder design requires slightly higher memory than a MWPM decoder 
for low code distances; and outperforms the MWPM for larger distances, 
making it more scalable. 
Given that we only consider average weight fault configurations and not
the worst case for the MWPM decoder, we believe that our lower bound 
on the MWPM memory capacity is very optimistic and the UF decoder
actually surpasses the MWPM decoder in term of memory capacity 
much before distance 20 as observed in the figure.

Other decoders are much more memory intensive than our design.
The LUT decoder requires the storage of more than $2^{1000}$ correction
bit-strings for $d=11$ and ML decoders cost several MBs to GBs of memory 
depending on implementation and the code distance.

\section{Resource optimization} \label{sec:resource_opt}

For a system with large number $L$ of logical qubits, the most straightforward 
implementation allocates two decoders per logical qubits, one for each
type of error, $X$ and $Z$.
Thus, for the baseline design, the decoding logic uses $2L$ decoders. 
However, the utilization of each pipeline stage varies causing under-utilization of 
certain stages. Ideally, we want to reduce the number of decoders required 
for the overall system.
In this section, we optimize not only the total number of decoders 
but the exact number of module of each type Gr-Gen, DFS engine and Corr engine.

\begin{figure}[htb]
\centering
    \includegraphics[width=1.0\columnwidth]{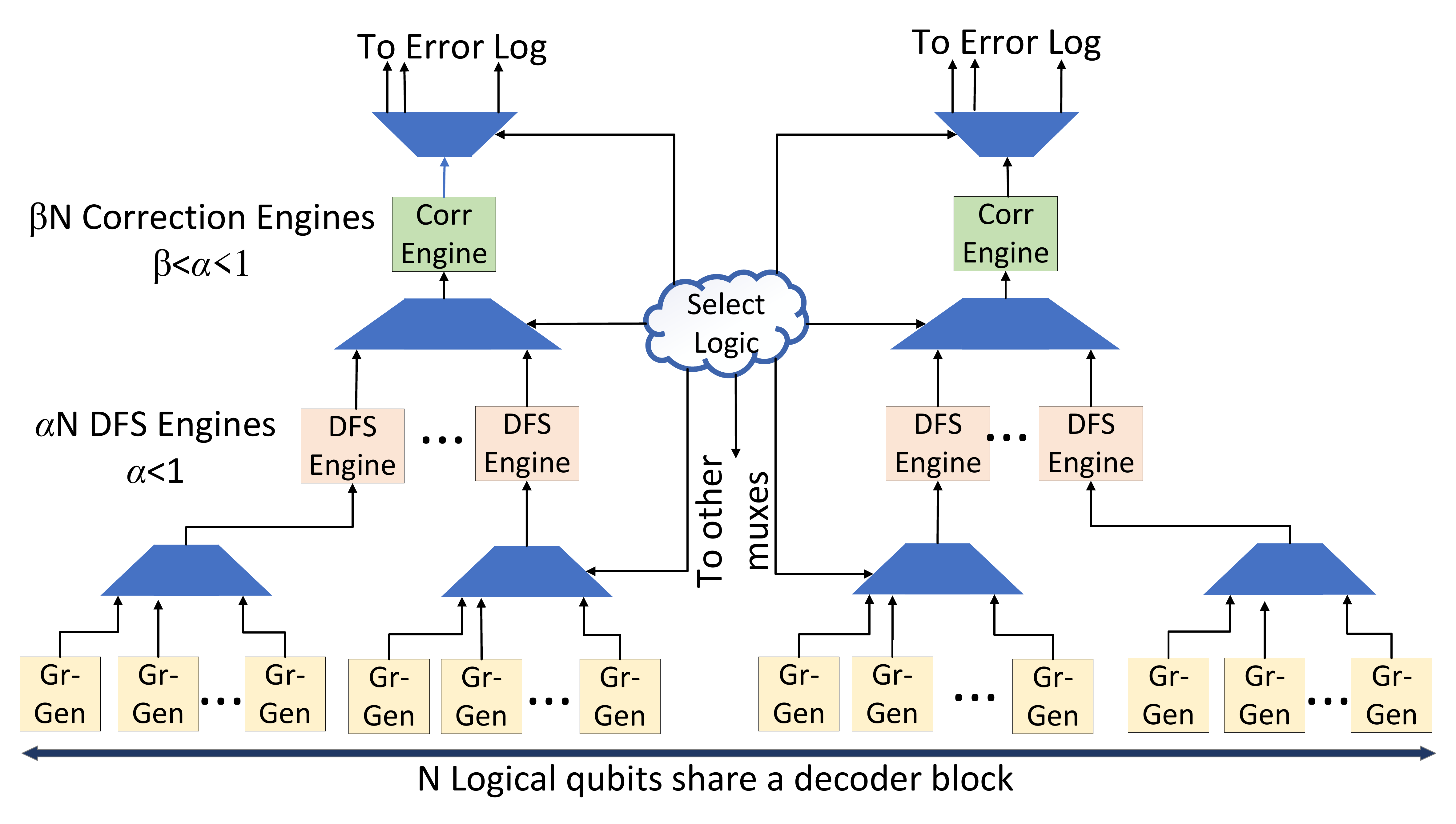}
    \caption{Design of decoder block that contains $N$ Gr-Gen modules,
    $\alpha N$ DFS engines and $\beta N$ Corr engines.}
    \label{fig:dblock}
\end{figure}

\subsection{Sharing hardware modules}

\nicolas{
The core component of our optimized architecture is a decoder block
as shown in Figure~\ref{fig:dblock} which uses a reduced number of 
pipeline units to perform the decoding of $N$ logical qubits.
Our {\em Error Decoding Architecture} (EDA) shown in Figure~\ref{fig:EDA}
uses $L/N$ decoder blocks to perform error correction over
the $L$ logical qubits of the computer.

The motivation for our design is that the growth stage implemented by the 
$N$ Gr-Gen modules is significantly more complex than the DFS stage. 
Therefore, we expect the DFS engine to wait for a fraction of the time while
the Gr-Gen module terminates.
Instead of waiting, we prefer to use a smaller number of DFS engines ($\alpha L < L$) 
that share the work of $L$ Gr-Gen modules. 
The value $\alpha$ will be optimized to keep the waiting time of DFS engines minimum,
saving a fraction ($1-\alpha$) of the DFS hardware.
We proceed in the same way to optimize the number 
$\beta L$ Corr engines.
}

The hardware overhead of this optimization is multiplexors and 
demultiplexors on the datapath as shown in Figure~\ref{fig:dblock}. 
Memory requests generated by the Corr Engine are routed to the 
correct memory locations using a demultiplexor. The select logic 
prioritizes the first ready component and uses round robin arbitration 
to generate appropriate select signals for the multiplexors. For example, 
if four Gr-Gen units share a DFS Engine, and the second Gr-Gen finishes 
cluster formation earlier than other units, it receives access to the 
DFS Engine. The round robin policy ensures fairness while sharing resources.

\subsection{Decode block design constraint}

\nicolas{
As explained previously, in order to correct circuit faults and measurement errors, 
the decoder needs $d$ consecutive rounds of syndrome data,
which form a {\em logical cycle}.
To prevent backlog problems, the decoder must provide a correction
before the end of the next logical cycle when a new decoding request arrives.
If a decoder block fails to terminate its work within a logical cycle, 
errors start accumulating and spreading over the quantum computer. 
We refer to this type of failure as \textit{timeout failure}. 
In order to ensure that timeout failures do not significantly degrade the 
decoding performance, we impose the following constraint for the decoder 
block design.
\begin{equation}
\label{eq:ptof}
p_{\Tof}(d, p) / N \leq p_{\Log}(d, p)
\end{equation}
where $p_{\Tof}(d, p)$ is the timeout failure probability for the decoder block.
The timeout failure probability per logical qubit must be lower than the probability 
of a logical error.
This condition ensures that the fault rate on the output of the whole quantum computation 
is at most doubled due to time out failures.
In what follows, we propose a fast and hardware efficient decoder block design 
that respects the constraint \eqref{eq:ptof}.
}

\subsection{Modules runtime simulation} \label{subsec:dblock}

We model the decoder performance by studying the number of reads. 
The write operations performed are read-modify-write, and the writeback 
is not on the critical path. We assume 4 cycles latency for memory accesses 
and a 4 GHz clock frequency~\cite{cacti,isca19cryo}. 

\nicolas{
Denote by $C_1, \dots C_m$ the set of clusters generated by the Gr-Gen
module.
A single growth step for a cluster $C$ requires to read a set of rows 
of the STM that cover the cluster.
We estimate this number by $\diam(C)^2$, assuming that the cluster spreads 
roughly uniformly in all the directions.
Summing of all clusters and growth steps, the total number of memory 
requests generated in Gr-Gen is approximated by the sum
\begin{equation} \label{eq:tref1}
\tau _{\GG} = \sum_{i=1}^m \sum_{j=1}^{\diam(C_i)} j^2
\end{equation}
because each growth steps increases the diameter by 1.

The number of memory requests in the DFS engine and the Corr engine
to treat a cluster $C_i$ are both given by the size of its spanning 
tree which is given by $|V(C_i)| - 1$ where $|V(C_i)|$ is the number 
of vertices of $C_i$.
Including all clusters, we obtain the estimate
\begin{equation} \label{eq:tref2}
\tau _{\DFS} = \tau _{\CE} = \sum_{i=1}^m |V(C_i)|
\end{equation}  
for the total number of reads in the DFS engine or in the Corr engine.
}

\begin{figure}[htb]
\centering
    \includegraphics[width=1.0\columnwidth]{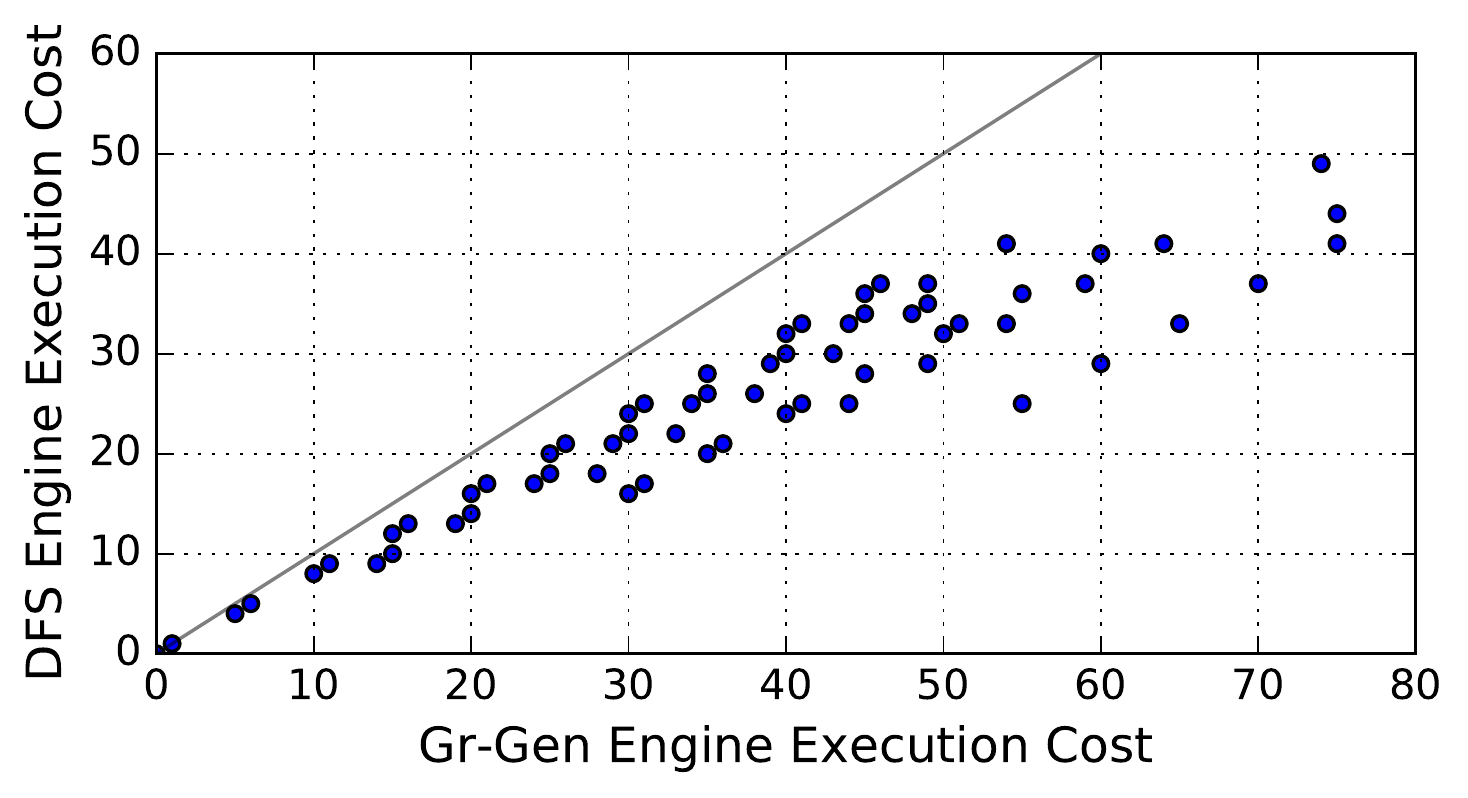}
    \caption{Correlation between Gr-Gen and DFS Engine execution time 
    for code distance $d=11$ and physical error rate $p=10^{-3}$. Each dot 
    corresponds to a random error configuration and the runtimes of the two modules
    are estimated using Eq.~\eqref{eq:tref1} and \eqref{eq:tref2}. Duplicate data points 
    cannot be observed on this plot.}
    \label{fig:correlation}
\end{figure}

In order to select the DFS engine ratio $\alpha$, 
we estimate the ratio between the execution time of the Gr-Gen and the
DFS engine by a Monte-Carlo simulation of $\tau _{\GG}$ and $\tau _{\DFS}$
for the distance-11 surface code with an error rate $p=10^{-3}$
as shown in Figure~\ref{fig:correlation}.
To estimate these runtimes, we sample families of clusters by generating 
random errors according 
to the phenomenological noise model described in 
Section~\ref{subsec:decoding_problem}, and by simulating the growth step 
of the decoder for these errors. This provides us with samples of cluster families
from the output of the Gr-Gen module.
Our Monte-Carlo simulation produces the result of Figure~\ref{fig:correlation}
which shows the correlation between the execution times in the Gr-Gen and DFS
engine. As expected more time is spent in the Gr-Gen unit. 
We observe roughly a factor two between the execution times of the two 
units which suggest one can eliminate half of the DFS units.

\subsection{Optimized decoder block} 

\begin{figure}[htb]
\centering
    \includegraphics[width=1.0\columnwidth]{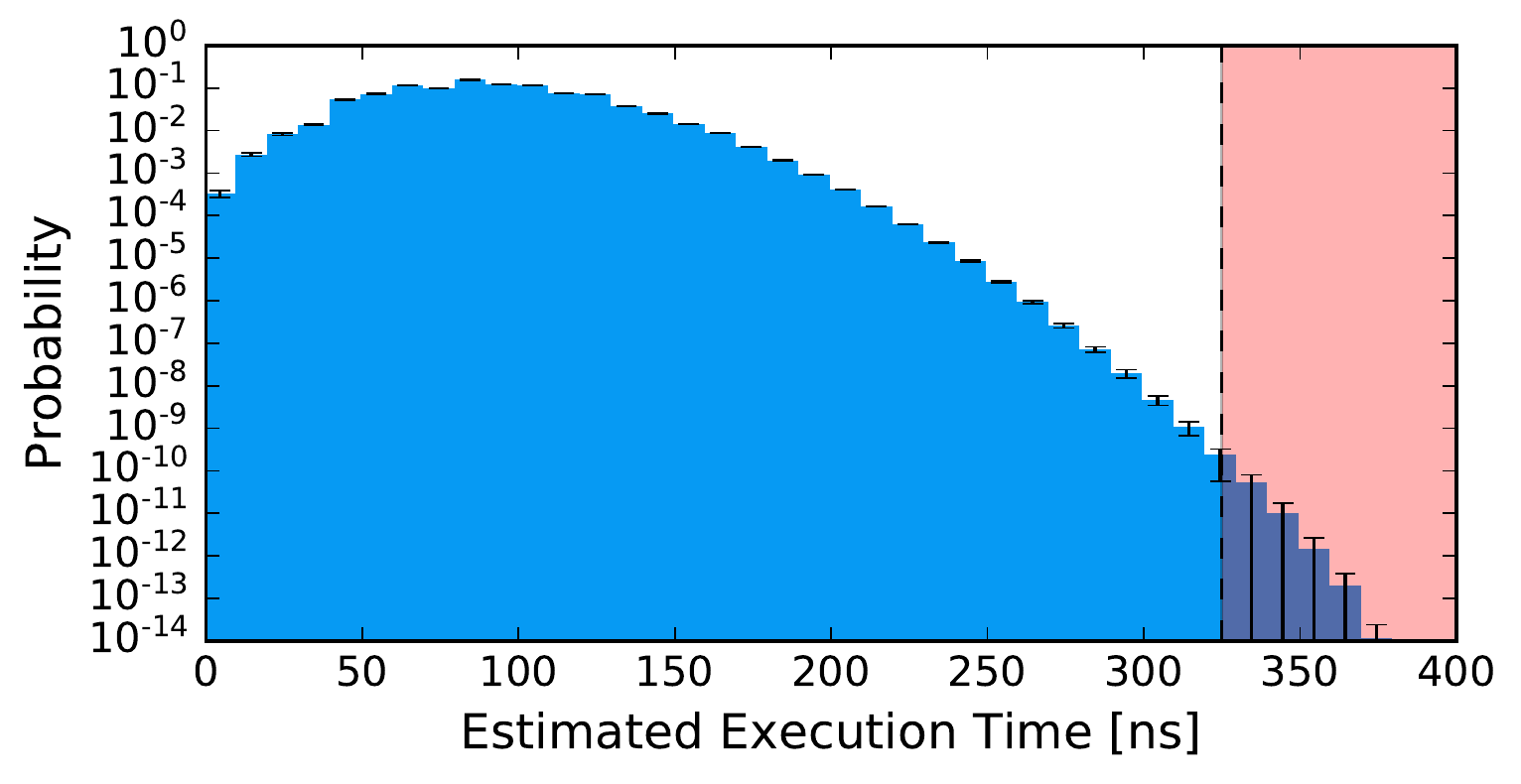}
    \caption{Distribution of execution time of the $(4, 2, 1, 1)$-block with code distance $11$ and error rate $10^{-3}$.
    The shaded region has probability smaller than $N p_{\Log}$ (for $N=2$ logical qubits), 
    which means that by interrupting the decoder block after 325 ns, we obtain a timeout failure that satisfies condition~\eqref{eq:ptof}.}
    \label{fig:exectime}
\end{figure}

\nicolas{
The results of Section~\ref{subsec:dblock} encourage us to consider 
a decoder block with parameters $\alpha = 0.5$ and $\beta = 1$.
However, nothing guarantees that this choice will lead to
a decoder block that is fast enough to satisfy condition~\eqref{eq:ptof}.
In this section, we design an optimized decoder block for the
surface code with distance 11 that satisfies~\eqref{eq:ptof} and that 
can be implemented in only 325ns in the noise regime $p=10^{-3}$,
under the memory frequency and latency assumptions above.
This demonstrates that our decoder block is clearly fast enough to perform
the surface code decoding. The decoder is actually 30 times faster that the 
logical cycle time of the distance-11 surface code which is about 11 $\mu s$
\cite{gidney2019RSA}.

We consider the smallest decoder block with $\alpha = 0.5$ and $\beta = 1$.
It includes two logical qubits, that is four error configurations to correct 
(two $X$-type and two $Z$-type), 
two Gr-Gen units, one DFS engine and one Corr engine.
We refer to this optimized design as the {\em $(4, 2, 1, 1)$-decoder block}.
Figure~\ref{fig:exectime} shows our estimation of the execution time of the 
whole block to decode the two logical qubits obtained by simulating the 
whole pipepline of the $(4, 2, 1, 1)$-block with a Monte-Carlo simulation
with importance sampling. 
We observe that by interrupting the decoding after 325 ns, we obtain a 
block that satisfy \eqref{eq:ptof}.
}

For $L$ logical qubits, the number of Gr-Gen units, DFS engines, and Corr engines required are L, L/2, and L/2 respectively in the optimized architecture. 
Thus, the total number of Gr-Gen units, DFS engines, and Corr engines are reduced by 2$\times$, 4$\times$,and 4$\times$ respectively.

The total memory capacity required for the baseline design and for our optimized
decoder block are summarized in Table~\ref{tab:hwreduction} for 
1000 qubits encoded with the distance-11 surface code.
The previous $(4,2,1,1)$-block leads to a saving of about $50\%$ of 
the memory capacity. In order to reduce further the memory requirement, 
we can use a shared root table and a shared size table between the two Gr-Gen 
modules of the decoder block. 
This leads to a slight slow down of the decoder because
both Gr-Gen modules cannot simultaneously perform the growth step, 
but the two STM can be used in parallel. While the first STM is used by a DFS engine, 
the second one can be used by a Gr-Gen module to grow clusters.
A $(4,2,1,1)$-block with a single root table and a single size table 
achieve $70\%$ (3.5 X) of memory reduction compare to the naive design.

\begin{table}[htb]
\begin{center}
\begin{footnotesize}
\caption{Reduction in the total memory capacity required to correct both $X$-type
and $Z$-type errors for a 1000 logical qubits with code distance 11 and 
error rate $10^{-3}$.}
\setlength{\tabcolsep}{1.2mm} 
\renewcommand{\arraystretch}{1.0}
\label{tab:hwreduction}
\begin{tabular}{ |c|c|c|c| } 
\hline
Design Component & Baseline & Optimized Design & Savings \\
\hline 
\hline
STM (Gr-Gen) & 1.97 MB & 0.99 MB & (2X) \\
\hline
Root Table (Gr-Gen) & 3.17 MB & 0.79 MB & (4X)  \\
\hline
Size Table (Gr-Gen) & 3.46 MB & 0.87 MB & (4X)  \\
\hline
Stacks (DFS Engine) & 1.35 MB & 0.34 MB & (4X) \\
\hline
Total & 9.96 & 2.81 & (3.5X) \\
\hline
\end{tabular}
\end{footnotesize}
\end{center}
\end{table}

\section{Syndrome data compression}
\label{sec:compression}

A major challenge in designing any error decoding architecture 
is the large bandwidth required between the quantum substrate and 
the decoding logic. 
In order to perform error decoding, the syndrome measurement data 
must be transported from the quantum substrate to the decoding units. 
For a given qubit plane with $L$ logical qubits and each qubit encoded 
using a surface code of distance $d$, about $2d^2 L$ bits must be sent 
at the end of each syndrome measurement cycle, which requires significant
bandwidth ranging on the order of several Gb/s for a reasonable number 
of logical qubits and code distance. 
Data transmission at a lower bandwidth on the other hand reduces 
the effective time left for error decoding since a decoder must provide 
an estimation of the error within $d$ syndrome measurement cycles. 
In this section, we consider three compression techniques to 
reduce the bandwidth needs and we analyze their performance in different 
noise regimes.
We focus on compression schemes that uses simple encoding 
and do not require large hardware complexity.

\subsection{Can data compression work?}

An approach to reduce the memory bandwidth requirements in conventional
memory systems is data compression~\cite{pekhimenko2013linearly,dzc,zhao2015buri,bitplanecompression,alameldeen2004adaptive,pekhimenko2012base}. Data compression works well when data is sparse and has low entropy. 
To justify the potential of syndrome data compression, we  
estimate the sparsity of the syndrome data analytically for realistic noise regimes.

\nicolas{
For a noise strength $p$ in the phenomenological noise model introduced
in Section~\ref{subsec:decoding_problem}, the expected number of faults
in the 3d decoding graph which contains $7d^3$ fault locations is about
$3 d^3 p$.
In the case of a distance-11 surface code, we expect only 4 errors 
on average, resulting in a non-trivial syndrome vector
of length $\approx 1,000$ with Hamming weight $\leq 8$,
since each error is detected by at most two non-trivial syndrome bits.
The expected Hamming weight of the syndrome vectors drops 
even further for lower noise strength $p$.
}

\begin{figure}[!tb]
\centering
    \includegraphics[width=1.0\columnwidth]{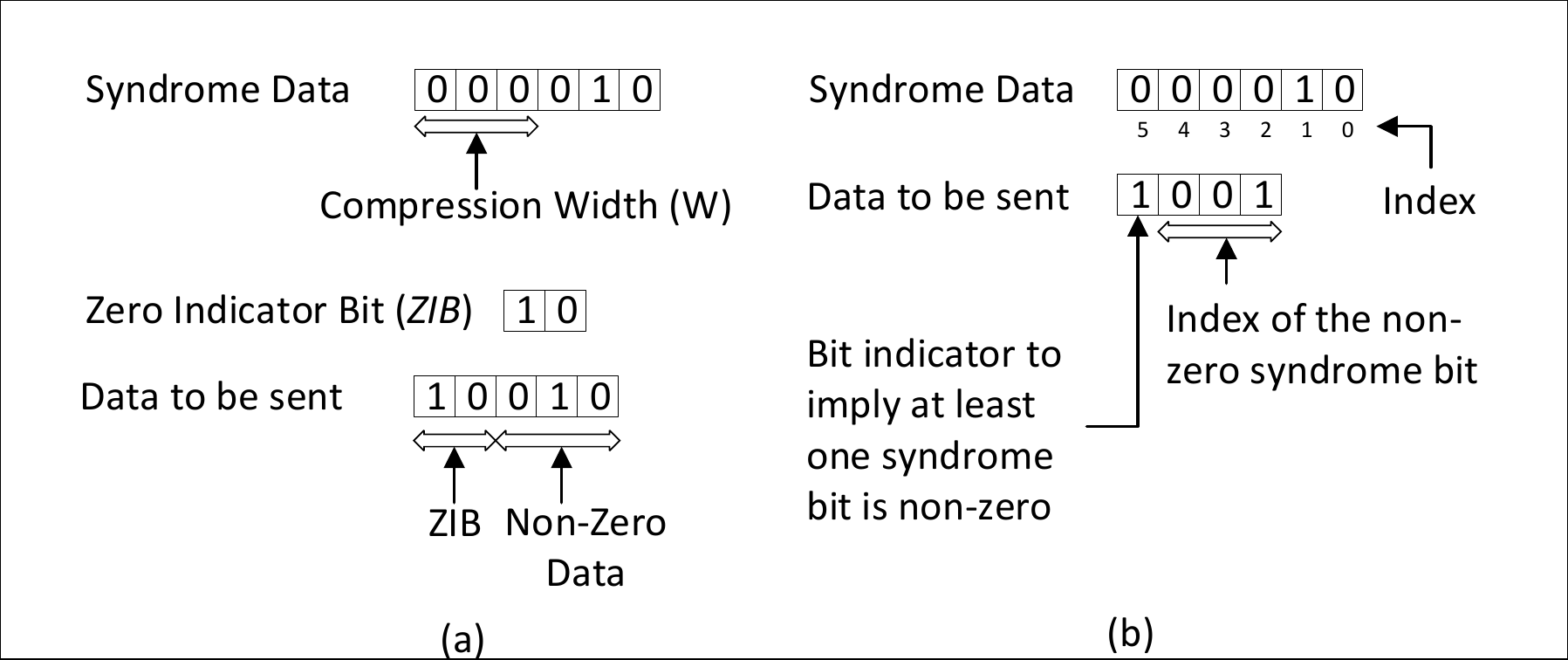}
    \caption{
    (a) Dynamic Zero Compression (DZC). The compressed data consists in
    the zero bit indicators, that provides the locations of blocks '000', followed
    by the content of non-zero blocks. 
     (b) Sparse Representation. We can store sparse efficiently data by 
     providing the number of non-zero bits and their index.
     }
    \label{fig:compression}
\end{figure}

\subsection{Three low-overhead compression techniques}

We consider the three following compression methods that allow
for a simple hardware implementation. Figure~\ref{fig:compression}
provides the basic idea of the Sparse Representation and the 
Dynamic Zero Compression. The Geometry-based compression 
is a variant of the Dynamic Zero Compression that exploits the 
geometry of the lattice of qubits.

\begin{enumerate}
    \item {\bf Sparse Representation}:
    This is similar to the traditional technique of storing only the indices 
    of non-zero elements of sparse matrices. 
    Instead of sending a sparse syndrome vectors, we send a bit that 
    indicates if all the syndrome bits are zero, followed by the indices of
    non-trivial bits in the case of non-zero syndrome.
    With this method, a syndrome vector with length $\ell$ and 
    Hamming weight $w$ is compressed into $1 + w \log_2(\ell)$ bits
    
    \item {\bf Dynamic Zero Compression (DZC)}: 
    We adopt a DZC technique~\cite{dzc} as shown in Figure~\ref{fig:compression}(a) 
    to compress syndrome data. 
    A syndrome vector of length $\ell$ is grouped into $b$ blocks of $m$ bits each.
    We store the indicator vector of non-trivial blocks concatenated with 
    the exact value of non-trivial blocks.
    A syndrome with $b$ block has length $b m$. However, if it contains only $w$ non-trivial 
    blocks, it can be transmitted with the DZC technique using only $b$ bits for the
    non-trivial block indicator vector and $w m$ bits to send the non-trivial blocks, 
    that is a total of $b + w m$ bits.
   
    \item {\bf Geometry-based compression (Geo-Comp)}: 
    This is an adaptation of DZC that also accounts for the geometry of the surface 
    code lattice. The basic idea is that non-trivial syndrome values generally appear
    by pairs of neighbor bits. We can therefore increase the compression rate of the 
    DZC technique by using square blocks that respect the structure of the lattice.
    With this block decomposition, two neighbors bits are more likely to fall in the same
    block, reducing the number $w$ of non-trivial blocks to send.
\end{enumerate}

\nicolas{
In this work, we treat $X$-type and $Z$-type errors separately.
However, for any of the three compression techniques described above, 
a slightly better compression rate can be obtained by compressing 
together the syndrome data corresponding to both types of errors.
}

In general, the number and size of the ZDC blocks can be adjusted for a 
given noise model by computing the expected number of non-trivial syndrome blocks.
Regions of larger size improves the compression ratio but also leads to complex 
hardware by adding to the logic depth. 
With this in mind, we analyze small block sizes even for very low error rates.

\begin{figure}[htb]
\centering
    \includegraphics[width=1.0\columnwidth]{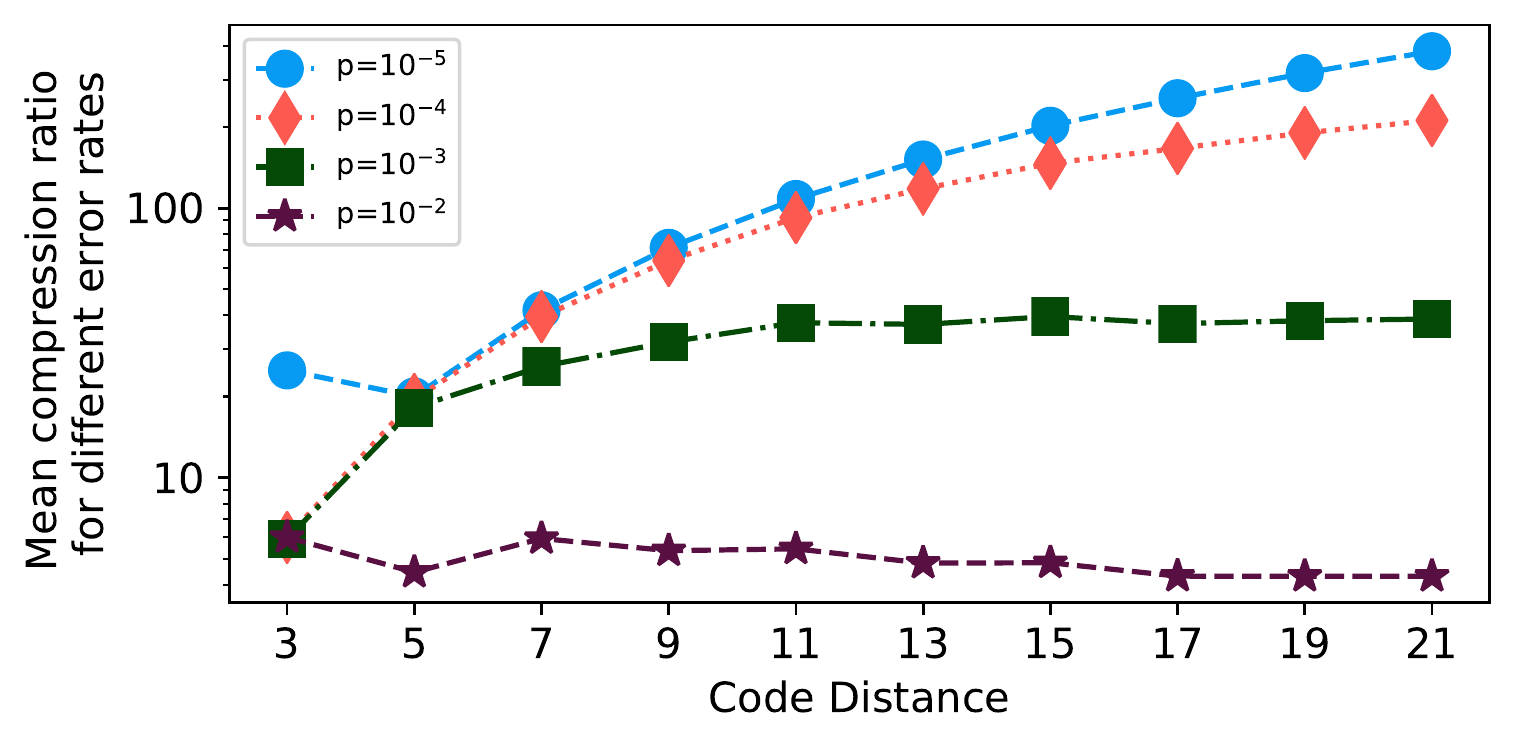}
    \caption{Mean syndrome compression ratio as a function of the code distance $d$ 
    for different physical error rates. We plot the best compression ratio between the three techniques
    considered in the present work: Sparse Representation, ZDC and Geometry-based compression.}
    \label{fig:compratio}
\end{figure} 

We determine the effectiveness of a compression scheme by analyzing 
the compression ratio:
\begin{equation} \label{eq:compratio}
\textrm{Compression Ratio} = \frac{\textrm{Actual Syndrome Length}}{\textrm{Compressed Syndrome Length}}
\end{equation}
The best compression scheme depends on the noise model. 
Sparse compression appears to be a good choice in the regime of
very low error rates because the syndrome vector is often trivial
and it is then sent as a single bit.
In a noisy regime and for small distance codes, we prefer the other 
compression schemes like DZC and Geometry-based compression.
For any noise rate below $10^{-3}$, we achieve a compression rate
which varies between 20$\times$ for $d=3$ and $400 \times$
for distance-21 surface codes.

\section{Discussion}
\subsection{Scalability}
In this paper, we use the Union-Find decoder to analyse the scope of architectural optimizations in designing high performance and scalable decoders. However, the design principles apply to other decoders in general such as machine learning or graph algorithm-based decoders. For example, machine learning decoders require several network layers.
Overall resources can be reduced by pipelining layers in inference and by sharing these layers between logical qubits. Similarly, while our study only focuses on regular surface code, the same analysis holds true for other types of QEC codes based on Euclidean lattices~\cite{fujii2012error, delfosse2016GSC} or color codes \cite{bombin2006color_codes}. However, it seems non-trivial to adapt the STM used in our design to the non-trivial lattice topology of hyperbolic codes \cite{zemor2009cayley, delfosse2013tradeoffs, breuckmann2016constructions} and thus a different micro-architecture is needed. Lastly, the syndrome compression is valid for an arbitrary decoder-quantum substrate interface, independent of their types, although it can be improved by exploiting the code structure as we propose with the geometry based compression.

\subsection{Assumptions}
We assume all syndrome extraction circuits can be executed in parallel, an assumption used by most decoders~\cite{versluis2017scalable}. However, the amount of parallelism depends on the qubit technology and a large amount of parallelism is achievable on modern superconducting qubits, although other types of qubits may offer less parallelism. 

\subsection{Noise Model}
We use the phenomenological noise model for our study and there is scope to further optimize the design for enhanced noise models and account for correlation in errors. We consider a physical error rate of $10^{-3}$ because QEC codes cannot lower the logical error rate substantially unless the initial physical error rate is lower than the threshold which is about 1\%. For the system sizes we have considered in this paper with about $100-1000$ qubits, error rates of about $10^{-3}$ are required to run practical applications of scientific and commercial value. 

\section{Related Work}
\label{sec:relatedwork}

Designing a quantum computer requires full-stack solutions~\cite{chong2017programming} and interdisciplinary research~\cite{martonosi2019next}. This has led to developments in programming languages~
QPL
\cite{
wecker2014liquid,
svore2018q,
websiteQsharp,
cross2017open,
websiteQiskit,
smith2016practical,
websitepyQuil,
websiteCirq,
websiteQasm,
khammassi2018cqasm,
green2013quipper,
websiteQuipper,
abhari2012scaffold,
websiteScaffold,
paykin2017qwire,
websiteQWIRE,
mccaskey2019xacc,
websiteQCOR,
bergholm2018pennylane,
websitePennylane,
killoran2019strawberryFields,
websiteStrawberryFields,
dahlberg2018simulaqron,
websiteSimulaCron,
steiger2018projectq,
websiteProjectQ}
compilers~\cite{qiskit,abhari2012scaffold,noiseadaptive,tannu2018ax,sabre,zulehner2018efficient,gokhale2019partial,heckey2015compiler}, micro-architecture~\cite{ding2018magic,tannu2017taming,fu2017experimental,javadi2017optimized}, 
control circuits~\cite{reilly2019challenges_crycontrol, mcdermott2014accurate, mcdermott2018quantum, li2019hardware, bardin2019cryocontrol, pauka2019cryocontrol}, and quantum devices. Although existing \textit{Noisy Intermediate Scale Quantum (NISQ)} computers~\cite{googlesupremacy,
hsu2018ces,
ibm53qubit,
ionq79,
aqt} 
are expected to scale up to hundreds of qubits and may outperform classical computers for some problems~\cite{vqe,qaoa,orus2019quantum}, they will still be too small to achieve fault-tolerance. On the contrary, the scope of a quantum computer greatly broadens in the presence of fault-tolerance and therefore, designing FTQCs is an important area of research. 

QEC plays a seminal role in FTQCs. 
In addition to the standard~\cite{preskill1998lecture, nielsen2002quantum}, 
we suggest the following recent reviews to learn more about 
QEC codes and fault tolerance~\cite{devitt2013quantum, terhal2015quantum, campbell2017roads}.
Recent experiment results suggest that quantum error correction is 
reaching an inflection point where the quality of encoded qubits is better 
than the quality of raw qubits \cite{vuillot2017QEC_experiment, harper2019QEC_experiment}.
Few articles consider the hardware aspects of decoder designs
are necessary such as discussing the potential of 
GPUs and ASICs~\cite{fowler2015parallel_MWPM, breuckmann2018NN_high_dimension_codes}, 
and high-speed circuits~\cite{fowler2012MWPM_timing} or describing 
the architecture of the neural networks in the case of ML-Decoders. 
These studies focus primarily on achieving higher performance and 
accuracy for a single logical qubit. 
In order to support the design of large scale error-corrected quantum
devices, more studies on the hardware aspects of decoder designs
and their scalability to many logical blocks are necessary.

In a system level study~\cite{tannu2017taming}, Tannu et. al. identified the requirement of large bandwidth in sending control instructions from the control processor to qubits and proposed sharing of micro-code between neighboring qubits. Using micro-code to deliver control pulses~\cite{fu2017experimental} only focuses on communication from the control processor to the qubits whereas, availability of large bandwidth is also essential to transmit syndrome data back from the qubits to the decoders. Since syndromes differ across logical qubits, depending on the error, it is not possible to send one syndrome for multiple qubits (sharing). So, we explore the possibility of using syndrome compression through low-overhead compression schemes. 

\section{Conclusion}
\label{sec:conclusion}
The decoder is a key component of a fault tolerant quantum computer, which is in charge of translating the output of the syndrome measurement circuits into error types and locations.
Many decoding algorithms have been studied in the past 20 years~\cite{
dennis2002tqm,
fowler2012autotune,
fowler2012MWPM,
fowler2012MWPM_timing,
fowler2015parallel_MWPM,
duclos2013qudit_RG_decoder,
anwar2014qudit_SC_decoder,
watson2015qudit_decoder,
hutter2015qudit_improved_HDRG,
duclos2010RG_decoder,
bravyi2013RG_cubic_code,
duclos2013RG_improved,
barrett2010loss_correction_short,
stace2010loss_correction_long,
delfosse2017peeling_decoder,
fowler2013XZ_correlations,
delfosse2014XZ_correlations,
criger2018XZ_correlations_and_degen,
tomita2014LUT_decoder,
heim2016optimal_decoder,
dennis2005phd,
harrington2004phd,
wootton2012high_threshold_decoder,
hutter2014MCMC_decoder,
wootton2015simple_decoder,
herold2015FT_SC_decoder,
breuckmann2016local_decoder_2d_4d,
delfosse2017UF_decoder,
kubica2019CA_decoder,
nickerson2019correlation_adaptive_decoder,
ferris2014TN_QEC,
bravyi2014MLD,
darmawan2018Lineartime_TN_decoding,
tuckett2018TN_biased,
torlai2017NN_SC_decoder,
baireuther2018NN_SC_correlations,
krastanov2017DNN_SC_decoder,
varsamopoulos2017NN_SC_decoding,
chamberland2018DNN_small_codes,
maskara2019NN_CC_decoder,
breuckmann2018NN_high_dimension_codes,
sweke2018RL_SC_decoder,
baireuther2019NN_CC_circuit_decoder,
ni2018NN_SC_large_d,
andreasson2019DNN_toric_code,
davaasuren2018CNN_SC_decoder,
liu2019NN_LDPC_codes,
varsamopoulos2018NN_SC_decoder_design,
varsamopoulos2019NN_SC_decoder_distributed,
wagner2019NN_training_with_symmetry,
chinni2019NN_CC_decoder,
sheth2019NN_SC_combining_decoder,
colomer2019RL_TC_decoder}.
They generally focus on improving either the decoder accuracy or the decoder runtime,
or providing a good compromise between them.
In this work, we introduce a third constraint, the scalability constraint, which
states that the decoder must scale to the regime of practical applications for which 
thousand of logical qubits must be decoded simultaneously
and we design a decoder that satisfy the three design constraints: accuracy, latency and scalability. 
Namely, the decoder properly identifies the error which occurs (accuracy), it is fast enough to avoid accumulation of errors during the computation (latency) and we propose an resource-efficient hardware implementation that scales to the massive size required for industrial applications (scalability).

In order to achieve the scalability constraint, we study the scope and impact of micro-architectural optimizations in designing decoders for QEC and study in-depth the Union-Find decoder as a case-study. We also investigate a system level framework for Error Decoding Architecture whereby instead of using dedicated decoding units per logical unit, we multiplex the decoding resources across neighbouring logical qubits while minimizing the timeout errors due to lack of decoding resources and limiting the possibility of system failure. Finally, we investigate the feasibility of low-cost \textit{syndrome compression} to reduce the memory bandwidth required for transmitting the syndrome information from the quantum substrate to the decoding hardware. 
Our solutions reduce the number of decoders by more than 50\%, the amount of memory required by 70\%, and memory bandwidth by more than 30x for large FTQCs. Although we use the Union-Find decoder and surface codes for our study, the design principles, optimizations, and results from our study applies to other types of decoders and certain other QEC codes as well. The compression schemes discussed applies to any qubit technology and decoder. 

In addition to substantial hardware savings, our optimized decoder micro-architecture significantly speeds up the Union-Find decoder. Our numerical simulations suggest that our design provides a decoder that is fast enough to perform error correction with superconducting qubits assuming a surface code syndrome round of $1 \mu s$ \cite{gidney2019RSA}. Further study is necessary to confirm the validity of our model in a real device. Ultimately, we would like to consider a FPGA implementation or the fabrication of an ASIC based on our micro-architecture.



\begin{thebibliography}{100}

\bibitem{shor1999polynomial}
Peter~W Shor.
\newblock Polynomial-time algorithms for prime factorization and discrete
  logarithms on a quantum computer.
\newblock {\em SIAM review}, 41(2):303--332, 1999.

\bibitem{feynman}
Richard~P Feynman.
\newblock Simulating physics with computers.
\newblock {\em International journal of theoretical physics}, 21(6):467--488,
  1982.

\bibitem{qchem}
Markus Reiher, Nathan Wiebe, Krysta~M Svore, Dave Wecker, and Matthias Troyer.
\newblock Elucidating reaction mechanisms on quantum computers.
\newblock {\em Proceedings of the National Academy of Sciences},
  114(29):7555--7560, 2017.

\bibitem{brown2010using}
Katherine~L Brown, William~J Munro, and Vivien~M Kendon.
\newblock Using quantum computers for quantum simulation.
\newblock {\em Entropy}, 12(11):2268--2307, 2010.

\bibitem{grover1996fast}
Lov~K Grover.
\newblock A fast quantum mechanical algorithm for database search.
\newblock {\em arXiv preprint quant-ph/9605043}, 1996.

\bibitem{shorqec}
Peter~W Shor.
\newblock Scheme for reducing decoherence in quantum computer memory.
\newblock {\em Physical review A}, 52(4):R2493, 1995.

\bibitem{calderbank1996good}
A~Robert Calderbank and Peter~W Shor.
\newblock Good quantum error-correcting codes exist.
\newblock {\em Physical Review A}, 54(2):1098, 1996.

\bibitem{steanecode}
Andrew Steane.
\newblock Multiple-particle interference and quantum error correction.
\newblock {\em Proceedings of the Royal Society of London. Series A:
  Mathematical, Physical and Engineering Sciences}, 452(1954):2551--2577, 1996.

\bibitem{gottesman1997stabilizer}
Daniel Gottesman.
\newblock Stabilizer codes and quantum error correction.
\newblock {\em arXiv preprint quant-ph/9705052}, 1997.

\bibitem{macwilliams1977ECC}
Florence~Jessie MacWilliams and Neil James~Alexander Sloane.
\newblock {\em The theory of error-correcting codes}, volume~16.
\newblock Elsevier, 1977.

\bibitem{shor1996ft}
Peter~W Shor.
\newblock Fault-tolerant quantum computation.
\newblock In {\em Proceedings of 37th Conference on Foundations of Computer
  Science}, pages 56--65. IEEE, 1996.

\bibitem{aharonov1999ft_threshold}
Dorit Aharonov and Michael Ben-Or.
\newblock Fault-tolerant quantum computation with constant error rate.
\newblock {\em arXiv preprint quant-ph/9906129}, 1999.

\bibitem{aliferis2005ft_threshold_d3}
Panos Aliferis, Daniel Gottesman, and John Preskill.
\newblock Quantum accuracy threshold for concatenated distance-3 codes.
\newblock {\em arXiv preprint quant-ph/0504218}, 2005.

\bibitem{dennis2002tqm}
Eric Dennis, Alexei Kitaev, Andrew Landahl, and John Preskill.
\newblock Topological quantum memory.
\newblock {\em Journal of Mathematical Physics}, 43(9):4452--4505, 2002.

\bibitem{fowler2012surface_code}
Austin~G Fowler, Matteo Mariantoni, John~M Martinis, and Andrew~N Cleland.
\newblock Surface codes: Towards practical large-scale quantum computation.
\newblock {\em Physical Review A}, 86(3):032324, 2012.

\bibitem{svore2007steane_threshold_2d}
Krysta~M Svore, David~P Divincenzo, and Barbara~M Terhal.
\newblock Noise threshold for a fault-tolerant two-dimensional lattice
  architecture.
\newblock {\em Quantum Information \& Computation}, 7(4):297--318, 2007.

\bibitem{fowler2012autotune}
Austin~G Fowler, Adam~C Whiteside, Angus~L McInnes, and Alimohammad Rabbani.
\newblock Topological code autotune.
\newblock {\em Physical Review X}, 2(4):041003, 2012.

\bibitem{fowler2012MWPM}
Austin~G Fowler, Adam~C Whiteside, and Lloyd~CL Hollenberg.
\newblock Towards practical classical processing for the surface code.
\newblock {\em Physical review letters}, 108(18):180501, 2012.

\bibitem{fowler2012MWPM_timing}
Austin~G Fowler, Adam~C Whiteside, and Lloyd~CL Hollenberg.
\newblock Towards practical classical processing for the surface code: Timing
  analysis.
\newblock {\em Physical Review A}, 86(4):042313, 2012.

\bibitem{fowler2015parallel_MWPM}
Austin~G Fowler.
\newblock Minimum weight perfect matching of fault-tolerant topological quantum
  error correction in average $ o (1) $ parallel time.
\newblock {\em Quantum Information and Computation}, 15(1\&2):0145--0158, 2015.

\bibitem{duclos2013qudit_RG_decoder}
Guillaume Duclos-Cianci and David Poulin.
\newblock Kitaev's z d-code threshold estimates.
\newblock {\em Physical Review A}, 87(6):062338, 2013.

\bibitem{anwar2014qudit_SC_decoder}
Hussain Anwar, Benjamin~J Brown, Earl~T Campbell, and Dan~E Browne.
\newblock Fast decoders for qudit topological codes.
\newblock {\em New Journal of Physics}, 16(6):063038, 2014.

\bibitem{watson2015qudit_decoder}
Fern~HE Watson, Hussain Anwar, and Dan~E Browne.
\newblock Fast fault-tolerant decoder for qubit and qudit surface codes.
\newblock {\em Physical Review A}, 92(3):032309, 2015.

\bibitem{hutter2015qudit_improved_HDRG}
Adrian Hutter, Daniel Loss, and James~R Wootton.
\newblock Improved hdrg decoders for qudit and non-abelian quantum error
  correction.
\newblock {\em New Journal of Physics}, 17(3):035017, 2015.

\bibitem{duclos2010RG_decoder}
Guillaume Duclos-Cianci and David Poulin.
\newblock Fast decoders for topological quantum codes.
\newblock {\em Physical review letters}, 104(5):050504, 2010.

\bibitem{bravyi2013RG_cubic_code}
Sergey Bravyi and Jeongwan Haah.
\newblock Quantum self-correction in the 3d cubic code model.
\newblock {\em Physical review letters}, 111(20):200501, 2013.

\bibitem{duclos2013RG_improved}
Guillaume Duclos-Cianci and David Poulin.
\newblock Fault-tolerant renormalization group decoder for abelian topological
  codes.
\newblock {\em arXiv preprint arXiv:1304.6100}, 2013.

\bibitem{barrett2010loss_correction_short}
Sean~D Barrett and Thomas~M Stace.
\newblock Fault tolerant quantum computation with very high threshold for loss
  errors.
\newblock {\em Physical review letters}, 105(20):200502, 2010.

\bibitem{stace2010loss_correction_long}
Thomas~M Stace and Sean~D Barrett.
\newblock Error correction and degeneracy in surface codes suffering loss.
\newblock {\em Physical Review A}, 81(2):022317, 2010.

\bibitem{delfosse2017peeling_decoder}
Nicolas Delfosse and Gilles Z{\'e}mor.
\newblock Linear-time maximum likelihood decoding of surface codes over the
  quantum erasure channel.
\newblock {\em arXiv preprint arXiv:1703.01517}, 2017.

\bibitem{fowler2013XZ_correlations}
Austin~G Fowler.
\newblock Optimal complexity correction of correlated errors in the surface
  code.
\newblock {\em arXiv preprint arXiv:1310.0863}, 2013.

\bibitem{delfosse2014XZ_correlations}
Nicolas Delfosse and Jean-Pierre Tillich.
\newblock A decoding algorithm for css codes using the x/z correlations.
\newblock In {\em 2014 IEEE International Symposium on Information Theory},
  pages 1071--1075. IEEE, 2014.

\bibitem{criger2018XZ_correlations_and_degen}
Ben Criger and Imran Ashraf.
\newblock Multi-path summation for decoding 2d topological codes.
\newblock {\em Quantum}, 2:102, 2018.

\bibitem{tomita2014LUT_decoder}
Yu~Tomita and Krysta~M Svore.
\newblock Low-distance surface codes under realistic quantum noise.
\newblock {\em Physical Review A}, 90(6):062320, 2014.

\bibitem{heim2016optimal_decoder}
Bettina Heim, Krysta~M Svore, and Matthew~B Hastings.
\newblock Optimal circuit-level decoding for surface codes.
\newblock {\em arXiv preprint arXiv:1609.06373}, 2016.

\bibitem{dennis2005phd}
Eric Dennis.
\newblock Purifying quantum states: Quantum and classical algorithms.
\newblock {\em arXiv preprint quant-ph/0503169}, 2005.

\bibitem{harrington2004phd}
James~William Harrington.
\newblock {\em Analysis of quantum error-correcting codes: symplectic lattice
  codes and toric codes}.
\newblock PhD thesis, California Institute of Technology, 2004.

\bibitem{wootton2012high_threshold_decoder}
James~R Wootton and Daniel Loss.
\newblock High threshold error correction for the surface code.
\newblock {\em Physical review letters}, 109(16):160503, 2012.

\bibitem{hutter2014MCMC_decoder}
Adrian Hutter, James~R Wootton, and Daniel Loss.
\newblock Efficient markov chain monte carlo algorithm for the surface code.
\newblock {\em Physical Review A}, 89(2):022326, 2014.

\bibitem{wootton2015simple_decoder}
James Wootton.
\newblock A simple decoder for topological codes.
\newblock {\em Entropy}, 17(4):1946--1957, 2015.

\bibitem{herold2015FT_SC_decoder}
Michael Herold, Michael~J Kastoryano, Earl~T Campbell, and Jens Eisert.
\newblock Fault tolerant dynamical decoders for topological quantum memories.
\newblock {\em arXiv preprint arXiv:1511.05579}, 2015.

\bibitem{breuckmann2016local_decoder_2d_4d}
Nikolas~P Breuckmann, Kasper Duivenvoorden, Dominik Michels, and Barbara~M
  Terhal.
\newblock Local decoders for the 2d and 4d toric code.
\newblock {\em arXiv preprint arXiv:1609.00510}, 2016.

\bibitem{delfosse2017UF_decoder}
Nicolas Delfosse and Naomi~H Nickerson.
\newblock Almost-linear time decoding algorithm for topological codes.
\newblock {\em arXiv preprint arXiv:1709.06218}, 2017.

\bibitem{kubica2019CA_decoder}
Aleksander Kubica and John Preskill.
\newblock Cellular-automaton decoders with provable thresholds for topological
  codes.
\newblock {\em Physical review letters}, 123(2):020501, 2019.

\bibitem{nickerson2019correlation_adaptive_decoder}
Naomi~H Nickerson and Benjamin~J Brown.
\newblock Analysing correlated noise on the surface code using adaptive
  decoding algorithms.
\newblock {\em Quantum}, 3:131, 2019.

\bibitem{ferris2014TN_QEC}
Andrew~J Ferris and David Poulin.
\newblock Tensor networks and quantum error correction.
\newblock {\em Physical review letters}, 113(3):030501, 2014.

\bibitem{bravyi2014MLD}
Sergey Bravyi, Martin Suchara, and Alexander Vargo.
\newblock Efficient algorithms for maximum likelihood decoding in the surface
  code.
\newblock {\em Physical Review A}, 90(3):032326, 2014.

\bibitem{darmawan2018Lineartime_TN_decoding}
Andrew~S Darmawan and David Poulin.
\newblock Linear-time general decoding algorithm for the surface code.
\newblock {\em Physical Review E}, 97(5):051302, 2018.

\bibitem{tuckett2018TN_biased}
David~K Tuckett, Christopher~T Chubb, Sergey Bravyi, Stephen~D Bartlett, and
  Steven~T Flammia.
\newblock Tailoring surface codes for highly biased noise.
\newblock {\em arXiv preprint arXiv:1812.08186}, 2018.

\bibitem{torlai2017NN_SC_decoder}
Giacomo Torlai and Roger~G Melko.
\newblock Neural decoder for topological codes.
\newblock {\em Physical review letters}, 119(3):030501, 2017.

\bibitem{baireuther2018NN_SC_correlations}
Paul Baireuther, Thomas~E O'Brien, Brian Tarasinski, and Carlo~WJ Beenakker.
\newblock Machine-learning-assisted correction of correlated qubit errors in a
  topological code.
\newblock {\em Quantum}, 2:48, 2018.

\bibitem{krastanov2017DNN_SC_decoder}
Stefan Krastanov and Liang Jiang.
\newblock Deep neural network probabilistic decoder for stabilizer codes.
\newblock {\em Scientific reports}, 7(1):11003, 2017.

\bibitem{varsamopoulos2017NN_SC_decoding}
Savvas Varsamopoulos, Ben Criger, and Koen Bertels.
\newblock Decoding small surface codes with feedforward neural networks.
\newblock {\em Quantum Science and Technology}, 3(1):015004, 2017.

\bibitem{chamberland2018DNN_small_codes}
Christopher Chamberland and Pooya Ronagh.
\newblock Deep neural decoders for near term fault-tolerant experiments.
\newblock {\em Quantum Science and Technology}, 3(4):044002, 2018.

\bibitem{maskara2019NN_CC_decoder}
Nishad Maskara, Aleksander Kubica, and Tomas Jochym-O'Connor.
\newblock Advantages of versatile neural-network decoding for topological
  codes.
\newblock {\em Physical Review A}, 99(5):052351, 2019.

\bibitem{breuckmann2018NN_high_dimension_codes}
Nikolas~P Breuckmann and Xiaotong Ni.
\newblock Scalable neural network decoders for higher dimensional quantum
  codes.
\newblock {\em Quantum}, 2:68--92, 2018.

\bibitem{sweke2018RL_SC_decoder}
Ryan Sweke, Markus~S Kesselring, Evert~PL van Nieuwenburg, and Jens Eisert.
\newblock Reinforcement learning decoders for fault-tolerant quantum
  computation.
\newblock {\em arXiv preprint arXiv:1810.07207}, 2018.

\bibitem{baireuther2019NN_CC_circuit_decoder}
Paul Baireuther, MD~Caio, B~Criger, Carlo~WJ Beenakker, and Thomas~E O’Brien.
\newblock Neural network decoder for topological color codes with circuit level
  noise.
\newblock {\em New Journal of Physics}, 21(1):013003, 2019.

\bibitem{ni2018NN_SC_large_d}
Xiaotong Ni.
\newblock Neural network decoders for large-distance 2d toric codes.
\newblock {\em arXiv preprint arXiv:1809.06640}, 2018.

\bibitem{andreasson2019DNN_toric_code}
Philip Andreasson, Joel Johansson, Simon Liljestrand, and Mats Granath.
\newblock Quantum error correction for the toric code using deep reinforcement
  learning.
\newblock {\em Quantum}, 3:183, 2019.

\bibitem{davaasuren2018CNN_SC_decoder}
Amarsanaa Davaasuren, Yasunari Suzuki, Keisuke Fujii, and Masato Koashi.
\newblock General framework for constructing fast and near-optimal
  machine-learning-based decoder of the topological stabilizer codes.
\newblock {\em arXiv preprint arXiv:1801.04377}, 2018.

\bibitem{liu2019NN_LDPC_codes}
Ye-Hua Liu and David Poulin.
\newblock Neural belief-propagation decoders for quantum error-correcting
  codes.
\newblock {\em Physical review letters}, 122(20):200501, 2019.

\bibitem{varsamopoulos2018NN_SC_decoder_design}
Savvas Varsamopoulos, Koen Bertels, and Carmen~G Almudever.
\newblock Designing neural network based decoders for surface codes.
\newblock {\em arXiv preprint arXiv:1811.12456}, 2018.

\bibitem{varsamopoulos2019NN_SC_decoder_distributed}
Savvas Varsamopoulos, Koen Bertels, and Carmen~G Almudever.
\newblock Decoding surface code with a distributed neural network based
  decoder.
\newblock {\em arXiv preprint arXiv:1901.10847}, 2019.

\bibitem{wagner2019NN_training_with_symmetry}
Thomas Wagner, Hermann Kampermann, and Dagmar Bru{\ss}.
\newblock Symmetries for a high level neural decoder on the toric code.
\newblock {\em arXiv preprint arXiv:1910.01662}, 2019.

\bibitem{chinni2019NN_CC_decoder}
Chaitanya Chinni, Abhishek Kulkarni, and Dheeraj~M Pai.
\newblock Neural decoder for topological codes using pseudo-inverse of parity
  check matrix.
\newblock {\em arXiv preprint arXiv:1901.07535}, 2019.

\bibitem{sheth2019NN_SC_combining_decoder}
Milap Sheth, Sara~Zafar Jafarzadeh, and Vlad Gheorghiu.
\newblock Neural ensemble decoding for topological quantum error-correcting
  codes.
\newblock {\em arXiv preprint arXiv:1905.02345}, 2019.

\bibitem{colomer2019RL_TC_decoder}
Laia~Domingo Colomer, Michalis Skotiniotis, and Ramon Muñoz-Tapia.
\newblock Reinforcement learning for optimal error correction of toric codes.
\newblock {\em arXiv preprint arXiv:1911.02308}, 2019.

\bibitem{fowler2017QEC_talk}
Austin Fowler.
\newblock Towards sufficiently fast quantum error correction.
\newblock Conference QEC 2017, 2017.

\bibitem{preskill1998lecture}
John Preskill.
\newblock Lecture notes for physics 229: Quantum information and computation.
\newblock {\em California Institute of Technology}, 16, 1998.

\bibitem{nielsen2002quantum}
Michael~A Nielsen and Isaac Chuang.
\newblock Quantum computation and quantum information, 2002.

\bibitem{fowler2013correlation}
Austin~G Fowler.
\newblock Optimal complexity correction of correlated errors in the surface
  code.
\newblock {\em arXiv preprint arXiv:1310.0863}, 2013.

\bibitem{delfosse2014XZcorrelations}
Nicolas Delfosse and Jean-Pierre Tillich.
\newblock A decoding algorithm for css codes using the x/z correlations.
\newblock In {\em 2014 IEEE International Symposium on Information Theory},
  pages 1071--1075. IEEE, 2014.

\bibitem{tuckett2018ultrahigh}
David~K Tuckett, Stephen~D Bartlett, and Steven~T Flammia.
\newblock Ultrahigh error threshold for surface codes with biased noise.
\newblock {\em Physical review letters}, 120(5):050505, 2018.

\bibitem{dennis2002topological}
Eric Dennis, Alexei Kitaev, Andrew Landahl, and John Preskill.
\newblock Topological quantum memory.
\newblock {\em Journal of Mathematical Physics}, 43(9):4452--4505, 2002.

\bibitem{raussendorf2007fault}
Robert Raussendorf and Jim Harrington.
\newblock Fault-tolerant quantum computation with high threshold in two
  dimensions.
\newblock {\em Physical review letters}, 98(19):190504, 2007.

\bibitem{raussendorf2007topological}
Robert Raussendorf, Jim Harrington, and Kovid Goyal.
\newblock Topological fault-tolerance in cluster state quantum computation.
\newblock {\em New Journal of Physics}, 9(6):199, 2007.

\bibitem{fowler2009highthreshold}
Austin~G Fowler, Ashley~M Stephens, and Peter Groszkowski.
\newblock High-threshold universal quantum computation on the surface code.
\newblock {\em Physical Review A}, 80(5):052312, 2009.

\bibitem{delfosse2017unionfind}
Nicolas Delfosse and Naomi~H Nickerson.
\newblock Almost-linear time decoding algorithm for topological codes.
\newblock {\em arXiv preprint arXiv:1709.06218}, 2017.

\bibitem{hsieh2011np}
Min-Hsiu Hsieh and Fran{\c{c}}ois Le~Gall.
\newblock Np-hardness of decoding quantum error-correction codes.
\newblock {\em Physical Review A}, 83(5):052331, 2011.

\bibitem{tomita2014low}
Yu~Tomita and Krysta~M Svore.
\newblock Low-distance surface codes under realistic quantum noise.
\newblock {\em Physical Review A}, 90(6):062320, 2014.

\bibitem{kolmogorov2009blossom}
Vladimir Kolmogorov.
\newblock Blossom v: a new implementation of a minimum cost perfect matching
  algorithm.
\newblock {\em Mathematical Programming Computation}, 1(1):43--67, 2009.

\bibitem{jouppi2017datacenter}
Norman~P Jouppi, Cliff Young, Nishant Patil, David Patterson, Gaurav Agrawal,
  Raminder Bajwa, Sarah Bates, Suresh Bhatia, Nan Boden, Al~Borchers, et~al.
\newblock In-datacenter performance analysis of a tensor processing unit.
\newblock In {\em 2017 ACM/IEEE 44th Annual International Symposium on Computer
  Architecture (ISCA)}, pages 1--12. IEEE, 2017.

\bibitem{delfosse2017peeling}
Nicolas Delfosse and Gilles Z{\'e}mor.
\newblock Linear-time maximum likelihood decoding of surface codes over the
  quantum erasure channel.
\newblock {\em arXiv preprint arXiv:1703.01517}, 2017.

\bibitem{tarjan1975UF}
Robert~Endre Tarjan.
\newblock Efficiency of a good but not linear set union algorithm.
\newblock {\em Journal of the ACM (JACM)}, 22(2):215--225, 1975.

\bibitem{cacti}
Naveen Muralimanohar, Rajeev Balasubramonian, and Norman~P Jouppi.
\newblock Cacti 6.0: A tool to model large caches.
\newblock {\em HP laboratories}, 27:28, 2009.

\bibitem{isca19cryo}
Gyu-hyeon Lee, Dongmoon Min, Ilkwon Byun, and Jangwoo Kim.
\newblock Cryogenic computer architecture modeling with memory-side case
  studies.
\newblock In {\em Proceedings of the 46th International Symposium on Computer
  Architecture}, pages 774--787. ACM, 2019.

\bibitem{gidney2019RSA}
Craig Gidney and Martin Eker{\aa}.
\newblock How to factor 2048 bit rsa integers in 8 hours using 20 million noisy
  qubits.
\newblock {\em arXiv preprint arXiv:1905.09749}, 2019.

\bibitem{pekhimenko2013linearly}
Gennady Pekhimenko, Vivek Seshadri, Yoongu Kim, Hongyi Xin, Onur Mutlu,
  Phillip~B Gibbons, Michael~A Kozuch, and Todd~C Mowry.
\newblock Linearly compressed pages: a low-complexity, low-latency main memory
  compression framework.
\newblock In {\em Proceedings of the 46th Annual IEEE/ACM International
  Symposium on Microarchitecture}, pages 172--184. ACM, 2013.

\bibitem{dzc}
Luis Villa, Michael Zhang, and Krste Asanovic.
\newblock Dynamic zero compression for cache energy reduction.
\newblock In {\em Proceedings 33rd Annual IEEE/ACM International Symposium on
  Microarchitecture. MICRO-33 2000}, pages 214--220. IEEE, 2000.

\bibitem{zhao2015buri}
Jishen Zhao, Sheng Li, Jichuan Chang, John~L Byrne, Laura~L Ramirez, Kevin Lim,
  Yuan Xie, and Paolo Faraboschi.
\newblock Buri: Scaling big-memory computing with hardware-based memory
  expansion.
\newblock {\em ACM Transactions on Architecture and Code Optimization (TACO)},
  12(3):31, 2015.

\bibitem{bitplanecompression}
Jungrae Kim, Michael Sullivan, Esha Choukse, and Mattan Erez.
\newblock Bit-plane compression: Transforming data for better compression in
  many-core architectures.
\newblock In {\em 2016 ACM/IEEE 43rd Annual International Symposium on Computer
  Architecture (ISCA)}, pages 329--340. IEEE, 2016.

\bibitem{alameldeen2004adaptive}
Alaa~R Alameldeen and David~A Wood.
\newblock Adaptive cache compression for high-performance processors.
\newblock {\em ACM SIGARCH Computer Architecture News}, 32(2):212, 2004.

\bibitem{pekhimenko2012base}
Gennady Pekhimenko, Vivek Seshadri, Onur Mutlu, Phillip~B Gibbons, Michael~A
  Kozuch, and Todd~C Mowry.
\newblock Base-delta-immediate compression: practical data compression for
  on-chip caches.
\newblock In {\em Proceedings of the 21st international conference on Parallel
  architectures and compilation techniques}, pages 377--388. ACM, 2012.

\bibitem{fujii2012error}
Keisuke Fujii and Yuuki Tokunaga.
\newblock Error and loss tolerances of surface codes with general lattice
  structures.
\newblock {\em Physical Review A}, 86(2):020303, 2012.

\bibitem{delfosse2016GSC}
Nicolas Delfosse, Pavithran Iyer, and David Poulin.
\newblock Generalized surface codes and packing of logical qubits.
\newblock {\em arXiv preprint arXiv:1606.07116}, 2016.

\bibitem{bombin2006color_codes}
Hector Bombin and Miguel~Angel Martin-Delgado.
\newblock Topological quantum distillation.
\newblock {\em Physical review letters}, 97(18):180501, 2006.

\bibitem{zemor2009cayley}
Gilles Z{\'e}mor.
\newblock On cayley graphs, surface codes, and the limits of homological coding
  for quantum error correction.
\newblock In {\em International Conference on Coding and Cryptology}, pages
  259--273. Springer, 2009.

\bibitem{delfosse2013tradeoffs}
Nicolas Delfosse.
\newblock Tradeoffs for reliable quantum information storage in surface codes
  and color codes.
\newblock In {\em 2013 IEEE International Symposium on Information Theory},
  pages 917--921. IEEE, 2013.

\bibitem{breuckmann2016constructions}
Nikolas~P Breuckmann and Barbara~M Terhal.
\newblock Constructions and noise threshold of hyperbolic surface codes.
\newblock {\em IEEE transactions on Information Theory}, 62(6):3731--3744,
  2016.

\bibitem{versluis2017scalable}
Richard Versluis, Stefano Poletto, Nader Khammassi, Brian Tarasinski, Nadia
  Haider, David~J Michalak, Alessandro Bruno, Koen Bertels, and Leonardo
  DiCarlo.
\newblock Scalable quantum circuit and control for a superconducting surface
  code.
\newblock {\em Physical Review Applied}, 8(3):034021, 2017.

\bibitem{chong2017programming}
Frederic~T Chong, Diana Franklin, and Margaret Martonosi.
\newblock Programming languages and compiler design for realistic quantum
  hardware.
\newblock {\em Nature}, 549(7671):180--187, 2017.

\bibitem{martonosi2019next}
Margaret Martonosi and Martin Roetteler.
\newblock Next steps in quantum computing: Computer science's role.
\newblock {\em arXiv preprint arXiv:1903.10541}, 2019.

\bibitem{wecker2014liquid}
Dave Wecker and Krysta~M Svore.
\newblock Liqui|>: A software design architecture and domain-specific language
  for quantum computing.
\newblock {\em arXiv preprint arXiv:1402.4467}, 2014.

\bibitem{svore2018q}
Krysta~M Svore, Alan Geller, Matthias Troyer, John Azariah, Christopher
  Granade, Bettina Heim, Vadym Kliuchnikov, Mariia Mykhailova, Andres Paz, and
  Martin Roetteler.
\newblock Q$\#$: Enabling scalable quantum computing and development with a
  high-level domain-specific language.
\newblock {\em arXiv preprint arXiv:1803.00652}, 2018.

\bibitem{websiteQsharp}
Microsoft.
\newblock Q$\#$, Accessed: November 19, 2019.
\newblock \url{https://docs.microsoft.com/en-us/quantum/?view=qsharp-preview}.

\bibitem{cross2017open}
Andrew~W Cross, Lev~S Bishop, John~A Smolin, and Jay~M Gambetta.
\newblock Open quantum assembly language.
\newblock {\em arXiv preprint arXiv:1707.03429}, 2017.

\bibitem{websiteQiskit}
IBM.
\newblock Qiskit, Accessed: November 19, 2019.
\newblock \url{https://qiskit.org/}.

\bibitem{smith2016practical}
Robert~S Smith, Michael~J Curtis, and William~J Zeng.
\newblock A practical quantum instruction set architecture.
\newblock {\em arXiv preprint arXiv:1608.03355}, 2016.

\bibitem{websitepyQuil}
Rigetti Computing.
\newblock pyquil, Accessed: November 19, 2019.
\newblock \url{http://docs.rigetti.com/en/stable/}.

\bibitem{websiteCirq}
Google.
\newblock Cirq, Accessed: November 19, 2019.
\newblock \url{https://github.com/quantumlib/Cirq}.

\bibitem{websiteQasm}
TU~Delft.
\newblock Qasm, Accessed: November 19, 2019.
\newblock \url{https://www.quantum-inspire.com/kbase/qasm/}.

\bibitem{khammassi2018cqasm}
N~Khammassi, GG~Guerreschi, I~Ashraf, JW~Hogaboam, CG~Almudever, and K~Bertels.
\newblock cqasm v1. 0: Towards a common quantum assembly language.
\newblock {\em arXiv preprint arXiv:1805.09607}, 2018.

\bibitem{green2013quipper}
Alexander~S Green, Peter~LeFanu Lumsdaine, Neil~J Ross, Peter Selinger, and
  Beno{\^\i}t Valiron.
\newblock Quipper: a scalable quantum programming language.
\newblock In {\em ACM SIGPLAN Notices}, volume~48, pages 333--342. ACM, 2013.

\bibitem{websiteQuipper}
Quipper, Accessed: November 19, 2019.
\newblock \url{https://www.mathstat.dal.ca/~selinger/quipper/}.

\bibitem{abhari2012scaffold}
Ali~J Abhari, Arvin Faruque, Mohammad~J Dousti, Lukas Svec, Oana Catu, Amlan
  Chakrabati, Chen-Fu Chiang, Seth Vanderwilt, John Black, and Fred Chong.
\newblock Scaffold: Quantum programming language.
\newblock Technical report, Princeton Univ. NJ Dept. of Computer Science, 2012.

\bibitem{websiteScaffold}
Scaffold, Accessed: November 19, 2019.
\newblock \url{https://scaffcc.llvm.org.cn/}.

\bibitem{paykin2017qwire}
Jennifer Paykin, Robert Rand, and Steve Zdancewic.
\newblock Qwire: a core language for quantum circuits.
\newblock In {\em ACM SIGPLAN Notices}, volume~52, pages 846--858. ACM, 2017.

\bibitem{websiteQWIRE}
Qwire, Accessed: November 19, 2019.
\newblock \url{https://github.com/inQWIRE/QWIRE}.

\bibitem{mccaskey2019xacc}
Alexander~J McCaskey, Dmitry~I Lyakh, Eugene~F Dumitrescu, Sarah~S Powers, and
  Travis~S Humble.
\newblock Xacc: A system-level software infrastructure for heterogeneous
  quantum-classical computing.
\newblock {\em arXiv preprint arXiv:1911.02452}, 2019.

\bibitem{websiteQCOR}
Qcor, Accessed: November 19, 2019.
\newblock \url{https://github.com/ORNL-QCI/qcor}.

\bibitem{bergholm2018pennylane}
Ville Bergholm, Josh Izaac, Maria Schuld, Christian Gogolin, Carsten Blank,
  Keri McKiernan, and Nathan Killoran.
\newblock Pennylane: Automatic differentiation of hybrid quantum-classical
  computations.
\newblock {\em arXiv preprint arXiv:1811.04968}, 2018.

\bibitem{websitePennylane}
Xanadu.
\newblock Pennylane, Accessed: November 19, 2019.
\newblock \url{https://pennylane.readthedocs.io/en/latest/}.

\bibitem{killoran2019strawberryFields}
Nathan Killoran, Josh Izaac, Nicol{\'a}s Quesada, Ville Bergholm, Matthew Amy,
  and Christian Weedbrook.
\newblock Strawberry fields: A software platform for photonic quantum
  computing.
\newblock {\em Quantum}, 3:129, 2019.

\bibitem{websiteStrawberryFields}
Xanadu.
\newblock Strawberry fields, Accessed: November 19, 2019.
\newblock \url{https://strawberryfields.readthedocs.io/en/latest/}.

\bibitem{dahlberg2018simulaqron}
Axel Dahlberg and Stephanie Wehner.
\newblock Simulaqron—a simulator for developing quantum internet software.
\newblock {\em Quantum Science and Technology}, 4(1):015001, 2018.

\bibitem{websiteSimulaCron}
Simulacron, Accessed: November 19, 2019.
\newblock \url{http://www.simulaqron.org/}.

\bibitem{steiger2018projectq}
Damian~S Steiger, Thomas H{\"a}ner, and Matthias Troyer.
\newblock Projectq: an open source software framework for quantum computing.
\newblock {\em Quantum}, 2(49):10--22331, 2018.

\bibitem{websiteProjectQ}
Projectq, Accessed: November 19, 2019.
\newblock \url{https://projectq.ch/}.

\bibitem{qiskit}
Andrew Cross.
\newblock The ibm q experience and qiskit open-source quantum computing
  software.
\newblock In {\em APS Meeting Abstracts}, 2018.

\bibitem{noiseadaptive}
Prakash Murali, Jonathan~M Baker, Ali~Javadi Abhari, Frederic~T Chong, and
  Margaret Martonosi.
\newblock Noise-adaptive compiler mappings for noisy intermediate-scale quantum
  computers.
\newblock {\em arXiv preprint arXiv:1901.11054}, 2019.

\bibitem{tannu2018ax}
Swamit~S Tannu and Moinuddin~K Qureshi.
\newblock A case for variability-aware policies for nisq-era quantum computers.
\newblock {\em arXiv preprint arXiv:1805.10224}, 2018.

\bibitem{sabre}
Gushu Li, Yufei Ding, and Yuan Xie.
\newblock Tackling the qubit mapping problem for nisq-era quantum devices.
\newblock {\em arXiv preprint arXiv:1809.02573}, 2018.

\bibitem{zulehner2018efficient}
Alwin Zulehner, Alexandru Paler, and Robert Wille.
\newblock Efficient mapping of quantum circuits to the ibm qx architectures.
\newblock In {\em 2018 Design, Automation \& Test in Europe Conference \&
  Exhibition (DATE)}, pages 1135--1138. IEEE, 2018.

\bibitem{gokhale2019partial}
Pranav Gokhale, Yongshan Ding, Thomas Propson, Christopher Winkler, Nelson
  Leung, Yunong Shi, David~I Schuster, Henry Hoffmann, and Frederic~T Chong.
\newblock Partial compilation of variational algorithms for noisy
  intermediate-scale quantum machines.
\newblock In {\em Proceedings of the 52nd Annual IEEE/ACM International
  Symposium on Microarchitecture}, pages 266--278. ACM, 2019.

\bibitem{heckey2015compiler}
Jeff Heckey, Shruti Patil, Ali JavadiAbhari, Adam Holmes, Daniel Kudrow,
  Kenneth~R Brown, Diana Franklin, Frederic~T Chong, and Margaret Martonosi.
\newblock Compiler management of communication and parallelism for quantum
  computation.
\newblock In {\em ACM SIGARCH Computer Architecture News}, volume~43, pages
  445--456. ACM, 2015.

\bibitem{ding2018magic}
Yongshan Ding, Adam Holmes, Ali Javadi-Abhari, Diana Franklin, Margaret
  Martonosi, and Frederic Chong.
\newblock Magic-state functional units: Mapping and scheduling multi-level
  distillation circuits for fault-tolerant quantum architectures.
\newblock In {\em 2018 51st Annual IEEE/ACM International Symposium on
  Microarchitecture (MICRO)}, pages 828--840. IEEE, 2018.

\bibitem{tannu2017taming}
Swamit~S Tannu, Zachary~A Myers, Prashant~J Nair, Douglas~M Carmean, and
  Moinuddin~K Qureshi.
\newblock Taming the instruction bandwidth of quantum computers via
  hardware-managed error correction.
\newblock In {\em 2017 50th Annual IEEE/ACM International Symposium on
  Microarchitecture (MICRO)}, pages 679--691. IEEE, 2017.

\bibitem{fu2017experimental}
Xiang Fu, MA~Rol, CC~Bultink, J~Van~Someren, Nader Khammassi, Imran Ashraf, RFL
  Vermeulen, JC~De~Sterke, WJ~Vlothuizen, RN~Schouten, et~al.
\newblock An experimental microarchitecture for a superconducting quantum
  processor.
\newblock In {\em Proceedings of the 50th Annual IEEE/ACM International
  Symposium on Microarchitecture}, pages 813--825. ACM, 2017.

\bibitem{javadi2017optimized}
Ali Javadi-Abhari, Pranav Gokhale, Adam Holmes, Diana Franklin, Kenneth~R
  Brown, Margaret Martonosi, and Frederic~T Chong.
\newblock Optimized surface code communication in superconducting quantum
  computers.
\newblock In {\em Proceedings of the 50th Annual IEEE/ACM International
  Symposium on Microarchitecture}, pages 692--705. ACM, 2017.

\bibitem{reilly2019challenges_crycontrol}
DJ~Reilly.
\newblock Challenges in scaling-up the control interface of a quantum computer.
\newblock {\em arXiv preprint arXiv:1912.05114}, 2019.

\bibitem{mcdermott2014accurate}
R~McDermott and MG~Vavilov.
\newblock Accurate qubit control with single flux quantum pulses.
\newblock {\em Physical Review Applied}, 2(1):014007, 2014.

\bibitem{mcdermott2018quantum}
R~McDermott, MG~Vavilov, BLT Plourde, FK~Wilhelm, PJ~Liebermann, OA~Mukhanov,
  and TA~Ohki.
\newblock Quantum--classical interface based on single flux quantum digital
  logic.
\newblock {\em Quantum science and technology}, 3(2):024004, 2018.

\bibitem{li2019hardware}
Kangbo Li, R~McDermott, and Maxim~G Vavilov.
\newblock Hardware-efficient qubit control with single-flux-quantum pulse
  sequences.
\newblock {\em Physical Review Applied}, 12(1):014044, 2019.

\bibitem{bardin2019cryocontrol}
Joseph~C Bardin, Evan Jeffrey, Erik Lucero, Trent Huang, Ofer Naaman, Rami
  Barends, Ted White, Marissa Giustina, Daniel Sank, Pedram Roushan, et~al.
\newblock 29.1 a 28nm bulk-cmos 4-to-8ghz!` 2mw cryogenic pulse modulator for
  scalable quantum computing.
\newblock In {\em 2019 IEEE International Solid-State Circuits
  Conference-(ISSCC)}, pages 456--458. IEEE, 2019.

\bibitem{pauka2019cryocontrol}
SJ~Pauka, K~Das, R~Kalra, A~Moini, Y~Yang, M~Trainer, A~Bousquet, C~Cantaloube,
  N~Dick, GC~Gardner, et~al.
\newblock A cryogenic interface for controlling many qubits.
\newblock {\em arXiv preprint arXiv:1912.01299}, 2019.

\bibitem{googlesupremacy}
Frank Arute, Kunal Arya, Ryan Babbush, Dave Bacon, Joseph~C Bardin, Rami
  Barends, Rupak Biswas, Sergio Boixo, Fernando~GSL Brandao, David~A Buell,
  et~al.
\newblock Quantum supremacy using a programmable superconducting processor.
\newblock {\em Nature}, 574(7779):505--510, 2019.

\bibitem{hsu2018ces}
Jeremy Hsu.
\newblock Ces 2018: Intels 49-qubit chip shoots for quantum supremacy.
\newblock {\em IEEE Spectrum Tech Talk}, 2018.

\bibitem{ibm53qubit}
The International Business~Machines Corporation.
\newblock Ibm raises the bar with a 50-qubit quantum computer.
\newblock {\em Sighted at Newsroom IBM:
  https://newsroom.ibm.com/2019-09-18-IBM-Opens-Quantum-Computation-Center-in-New-York-Brings-Worlds-Largest-Fleet-of-Quantum-Computing-Systems-Online-Unveils-New-53-Qubit-Quantum-System-for-Broad-Use},
  2019.

\bibitem{ionq79}
IonQ.
\newblock Ionq harnesses single-atom qubits to build the world’s most
  powerful quantum computer.
\newblock {\em Sighted at IonQ News: https://ionq.com/news/december-11-2018},
  2019.

\bibitem{aqt}
Alpine quantum technologies, Accessed: November 19, 2019.
\newblock \url{https://www.aqt.eu/}.

\bibitem{vqe}
Jarrod~R McClean, Jonathan Romero, Ryan Babbush, and Al{\'a}n Aspuru-Guzik.
\newblock The theory of variational hybrid quantum-classical algorithms.
\newblock {\em New Journal of Physics}, 18(2):023023, 2016.

\bibitem{qaoa}
Edward Farhi, Jeffrey Goldstone, and Sam Gutmann.
\newblock A quantum approximate optimization algorithm.
\newblock {\em arXiv preprint arXiv:1411.4028}, 2014.

\bibitem{orus2019quantum}
Roman Orus, Samuel Mugel, and Enrique Lizaso.
\newblock Quantum computing for finance: overview and prospects.
\newblock {\em Reviews in Physics}, page 100028, 2019.

\bibitem{devitt2013quantum}
Simon~J Devitt, William~J Munro, and Kae Nemoto.
\newblock Quantum error correction for beginners.
\newblock {\em Reports on Progress in Physics}, 76(7):076001, 2013.

\bibitem{terhal2015quantum}
Barbara~M Terhal.
\newblock Quantum error correction for quantum memories.
\newblock {\em Reviews of Modern Physics}, 87(2):307, 2015.

\bibitem{campbell2017roads}
Earl~T Campbell, Barbara~M Terhal, and Christophe Vuillot.
\newblock Roads towards fault-tolerant universal quantum computation.
\newblock {\em Nature}, 549(7671):172, 2017.

\bibitem{vuillot2017QEC_experiment}
Christophe Vuillot.
\newblock Is error detection helpful on ibm 5q chips?
\newblock {\em arXiv preprint arXiv:1705.08957}, 2017.

\bibitem{harper2019QEC_experiment}
Robin Harper and Steven~T Flammia.
\newblock Fault-tolerant logical gates in the ibm quantum experience.
\newblock {\em Physical review letters}, 122(8):080504, 2019.

\end{thebibliography}
\end{document}